\documentclass[twocolumn]{aastex63}
\usepackage{xcolor}
\usepackage{color}
\usepackage{graphicx}
\usepackage{dcolumn}
\usepackage{bm}
\usepackage{hyperref}
\usepackage{amsmath}
\usepackage{multirow}

\newcommand{\Ms}{{\rm ~M}_\odot}
\newcommand{\bh}{{\rm BH}}
\newcommand{\ns}{{\rm NS}}
\newcommand{\co}{{\rm CO}}
\newcommand{\gws}{{\rm GW190814}}
\newcommand{\yrgpc}{{\rm ~yr}^{-1}{\rm ~Gpc}^{-3}}
\newcommand{\lvk}{{\rm LVC}}
\newcommand{\gw}{{\rm GW}}
\newcommand{\der}{{\rm d}}

\received{}
\revised{}
\accepted{}
\submitjournal{ApJL}

\shorttitle{Dynamical formation of GW190814 merger}
\shortauthors{Arca Sedda, Manuel}
\graphicspath{{./}{}}

\begin{document}

\title{Dynamical formation of the GW190814 merger}

\correspondingauthor{Manuel Arca Sedda}
\email{m.arcasedda@gmail.com}

\author[0000-0002-3987-0519]{Manuel Arca Sedda}
\affil{Astronomisches Rechen-Institut, Zentrum f\"{u}r Astronomie der Universit\"{a}t  Heidelberg, M\"onchhofstr. 12-14, D-69120 Heidelberg, Germany}

\begin{abstract}
We investigate the possible dynamical origin of GW190814, a gravitational wave (GW) source discovered by the LIGO-Virgo-Kagra collaboration (LVC) associated with a merger between a stellar black hole (BH) with mass $23.2\Ms$ and a compact object, either a BH or a neutron star (NS), with mass $2.59\Ms$. Using a database of 240,000 $N$-body simulations modelling the formation of NS-BH mergers via dynamical encounters in dense clusters, we find that systems like GW190814 are likely to form in young, metal-rich clusters. Our model suggests that 
a little excess ($\sim 2-4\%$) of objects with masses in the range $2.3-3\Ms$ in the compact remnants' mass spectrum  leads to a detection rate for dynamically formed ``GW190814 -like'' mergers of $\Gamma_{\rm GW190814} \simeq 1-6\yrgpc$, i.e. within the observational constraints set by the GW190814 discovery, $\Gamma_\lvk \sim 1-23\yrgpc$. Additionally, our model suggests that $\sim 1.8-4.8\%$ of dynamical NS-BH mergers are compatible with GW190426\_152155, the only confirmed NS-BH merger detected by the LVC. We show that the relative amount of light and heavy NS-BH mergers can provide clues about the environments in which they developed.  
\end{abstract}

\section{Introduction}
\label{sec:intro} 
The LIGO-Virgo collaboration (LVC) detected recently GW190814, a merger between a BH with mass $M_\bh = 23.2_{-1.0}^{+1.1}\Ms$ and a \textit{mysterious} compact object with mass $M_\co = 2.59_{-0.09}^{+0.08} \Ms$ \citep{190814}. The properties of GW190814 challenge our understanding of compact binaries: i) the secondary mass falls in the ``lower mass gap'', a range of masses ($2.5-5\Ms$) characterised by the observational absence of stellar remnants \citep{Bailyn98,Ozel12}, ii) the mass ratio is small, $q = 0.112^{+0.008}_{-0.008}$, iii) and the inferred merger rate is fairly large, $\Gamma_\lvk = 7^{+16}_{-6}\yrgpc$.
The unusual mass of GW190814 secondary suggests that this merger involved either the heaviest NS or the lightest BH known in a compact binary system. Although the BH hypotesis seems to be the favoured one \citep{190814}, the existence of NS with masses up to $3\Ms$ \citep[e.g.][]{Freire08,ruiz20} or, more in general, the absence of a lower mass gap \citep[e.g.][]{Wyrzykowski20,zevin20} cannot be completely ruled out. Whether the secondary is a NS or a BH, matching all GW190814 features -- low-mass companion, low mass-ratio, and large merger rate -- poses a challenge to astrophysical theories. Population synthesis models for isolated binaries predict mass ratios $q>0.2$ \citep{dominik12,marchant17,giacobbo18,spera19}, unless special prescriptions are adopted \citep{Eldridge17,giacobbo18}. Formation in active galactic nuclei could be a promising channel, although mergers developing in such extreme environments might have larger BH masses, $M_\bh \sim 50\Ms$ \citep{yang20}, but comparable mass ratios, i.e. $q = 0.07-0.2$ \citep{yang20,mckernan20}, than GW190814. Nonetheless, $\sim 4\%$ of AGN-assisted mergers can have one of the binary components in the lower mass-gap \citep{yang20b}. Other explanations include the accretion of material expelled during the NS formation remained bound to the binary due to the large mass of the primary \citep{safarzadeh20}, through the development of a hierarchical merger involving two NSs and a BH \citep{Lu21} or, more in general, in hierarchical triples assembled either in the field or in dense clusters \citep[e.g.][]{liu20}. Alternatively, GW190814-like systems may hint at a supernova (SN) mechanism acting on longer timescales than previously thought, thus enabling the proto-compact remnant to accrete enough mass before undergoing explosion \citep{zevin20}. However, even in such case the formation of mergers with a secondary mass and mass ratio compatible with GW190814 is almost impossible in the isolated binary scenario, regardless of the SN explosion mechanism assumed \citep{zevin20}. Another potential formation channel for GW190814 is via dynamical encounters in a star cluster. The dynamical formation of massive binaries (e.g. BH-BH) is efficient in globular clusters (GCs), where compact remnants undergo dozens of interactions before either merging inside the cluster, or getting ejected and merge afterward \citep[][]{rodriguez16,askar17,rodriguez18}. 
However, binary BHs in GCs tend to have high mass ratios \citep[$q > 0.5$,][]{rodriguez16}, while the formation of NS-BH binaries is suppressed owing to BHs that quench mass segregation of lighter objects. Therefore, the inferred NS-BH merger rate for GCs in the local Universe is rather low, $\Gamma_{\rm NSBH} = 10^{-2}-10^{-1}$ \citep{clausen13,arcasedda20,Ye2020}. Young and open clusters (YCs) might be suitable formation sites for NS-BH mergers \citep{ziosi14,rastello20}. The large number of YCs expected to lurk in galaxies \citep[up to $10^5$ in Milky Way, e.g.][]{piskunov06} can boost the overall NS-BH merger rate, especially if they contain a high fraction of primordial binaries \citep{rastello20}. In a recent work, we explored the NS-BH dynamical formation channel through a suite of 240,000 $N$-body simulations \citep{arcasedda20} tailored to reproduce scatterings in star clusters with velocity dispersion $\sigma = 5-100$ km s$^{-1}$, thus covering the range from YCs to nuclear clusters (NCs). We predict that some NS-BH mergers display distictive features, namely a chirp mass $\mathcal > 4\Ms$, a BH heavier than $M_\bh > 10\Ms$, and the absence of an electromagnetic (EM) counterpart even if the BH is highly spinning. In this Letter, we exploit this database to quantify the likelihood for dynamically formed GW190814-like sources.

\section{Dynamical NS-BH merger rates and the formation of GW190814-like sources}
\label{sec:formation} 

\begin{figure}
    \centering
    \includegraphics[width=\columnwidth]{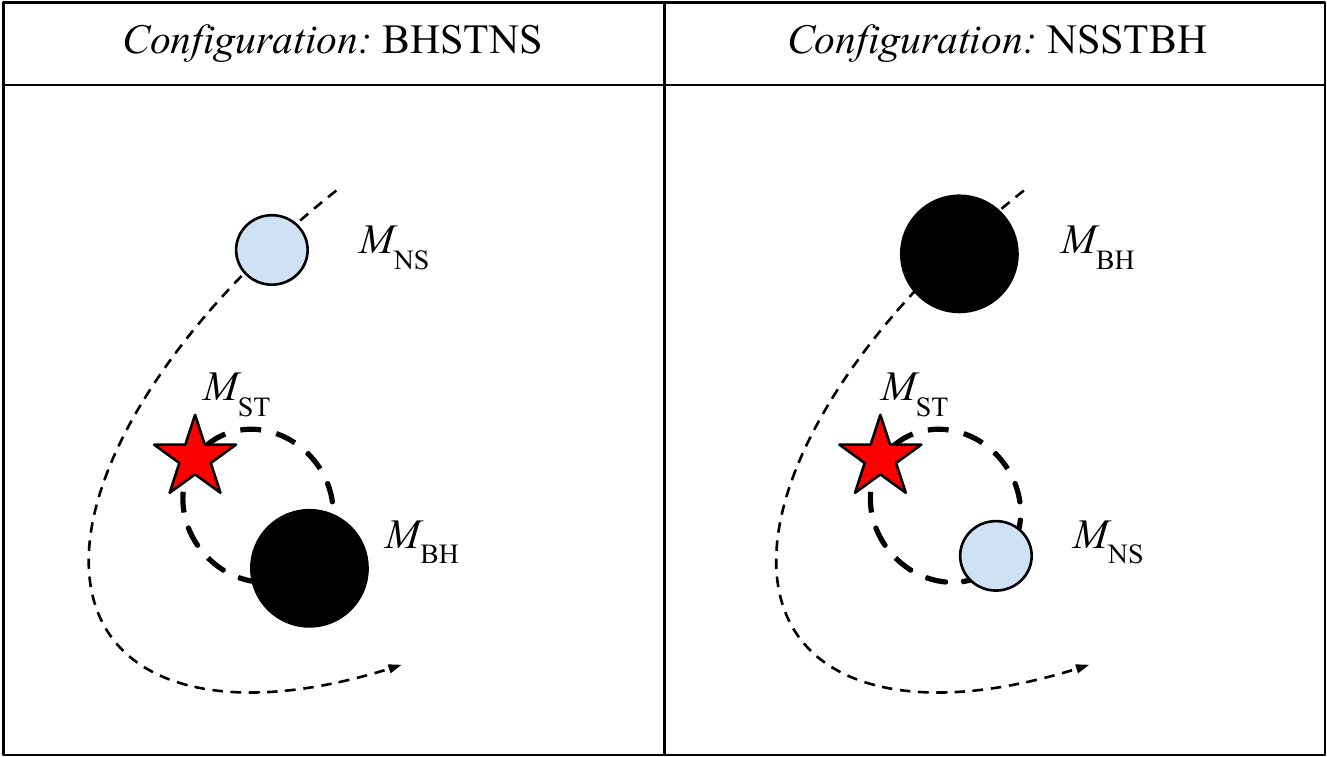}
    \caption{Schematization of the dynamical encounter driving the formation of a neutron star - black hole binary (NS-BH) in a dense stellar environment. We assume that either a roaming NS scatters over a BH-star (ST) binary (left panel, configuration BHSTNS) or vice-versa (right panel, NSSTBH).}
    \label{fig:sketch}
\end{figure}

Our simulations \citep{arcasedda20} model binary-single hyperbolic encounters involving two compact objects and a {\it normal} star through \texttt{ARGdf} \citep{arcasedda19}, an improved version of the \texttt{ARCHAIN} $N$-body code that implements Post-Newtonian formalism up to order 2.5 \citep{mikkola08} and enables a high-accuracy treatment for close encounters \citep{mikkola99}. We adopt a mass function for compact objects such that all remnants with a final mass $<3\Ms$ are labelled as NSs, whilst the remaining are labelled as BHs. The database, containing over 240,000 simulations, is dissected into two configurations (see Figure \ref{fig:sketch}): either the binary contains a BH and a star (ST) and the third object is a NS (configuration BHSTNS), or viceversa (NSSTBH). Details about the initial conditions and the mass distribution of the objects involved in the scattering are discussed in Appendix \ref{sec:app1} and \ref{sec:app2}.

We identify $N_{\rm GW} = 1193$ mergers, i.e. $P_{\ns-\bh} \simeq 0.5\%$ of the whole sample.
To characterise how $N_{\rm GW}$ varies across different values of the velocity dispersion ($\sigma$), we define an \textit{individual merger rate} ($\Gamma_{\rm ind}$), namely the number of mergers per unit time per cluster, as 
\begin{equation}
\Gamma_{\rm ind} = p_{\rm GW} N_{\rm bin} {\rm d} R/{\rm d} t,
\label{eq:0}
\end{equation}
i.e. as the product between the fraction of NS-BH mergers ($p_{\rm GW}$) which is measured directly from our simulations, the average number of binaries containing either a NS or a BH that at a given time coexist in the cluster ($N_{\rm bin}$), and the rate of binary-single interactions that lead to the formation of a NS-BH binary (${\rm d} R/{\rm d} t$). We find that there is a tight relation between $\Gamma_{\rm ind}$ and $\sigma$, well described by a power-law
\begin{equation}
    \Gamma_{\rm ind} \simeq k N_{\rm bin} \left(\frac{\sigma}{\sigma_c}\right)^\delta.
    \label{eq:1}
\end{equation}
In the equation above, $\sigma_c = 5$ km s$^{-1}$, and $k$ and $\delta$, whose values are summarized in Table \ref{tab:tab1}, are best-fit parameters calculated through a linear regression fit applied to the database of simulated mergers. The functional form above is likely the result of the relation between the parameters involved in Equation \ref{eq:0} and the cluster velocity dispersion. As discussed in Appendix \ref{sec:app1}, the $N_{\rm bin}$ parameter is highly uncertain, as it depends on the mass of the cluster, the retention fraction of both NSs and BHs, and the cluster relaxation time. To constrain this quantity we resort to the suite of GCs Monte Carlo models named the MOCCA Survey Database I \citep{askar17}. In MOCCA we find on average $N_{\rm bin} \sim 1$ for cluster mass $M<10^5\Ms$ and $N_{\rm bin} = 2-4$ for heavier clusters.
Note that the cluster mass $M$ and half-mass radius $r_h$ are linked to $\sigma$ via \citep{arcasedda20}
\begin{equation}
{\rm Log} \left(GM/r_h\right) = (1.14\pm0.03) +2{\rm Log}\sigma,
\label{eq:2}
\end{equation}
thus we can uniquely infer the individual merger rate for a cluster with given mass and half-mass radius through its velocity dispersion. Table \ref{tab:tab2} lists $\Gamma_{\rm ind}$ estimates assuming typical values for $\sigma$ in YCs, GCs, and NCs. At a redshift $z\lesssim 1$, the merger rate associated with a given cluster population can be roughly calculated as \citep{arcasedda20,Ye2020}:
\begin{equation}
    \Gamma_{\rm NS-BH} = \Gamma_{\rm ind} \rho_{\rm MWEG} N_c,
\end{equation}
where $\Gamma_{\rm ind}$ is calculated through Eq. \ref{eq:1}, $\rho_{\rm MWEG} = 0.0116$ Mpc$^{-3}$ is the local density of Milky Way equivalent galaxies \citep{abadie10} and $N_c$ is the total number of clusters in the galaxy. A typical NC has $\Gamma_{\rm NC} < 0.5$ Gyr$^{-1}$, i.e. 1-3 orders of magnitude larger than other cluster types. However, the contribution of NCs to the population of NS-BH is likely rather low, as they are outnumbered by GCs \citep[around 200 in the Milky Way][]{harris10} and YCs \citep[up to $10^5$][]{piskunov06}. Assuming around 200 GCs and 1 NCs for all MW-like galaxies in the local Universe implies a merger rate $\Gamma_{\rm GC} = (0.002-0.1) \times N_{\rm bin} \yrgpc$ and $\Gamma_{\rm NC} = (3\times 10^{-5} - 6\times 10^{-3}) \times N_{\rm bin} \yrgpc$. To explain the LVC inferred rate, the number of binaries with a compact object lurking in typical GCs and NCs should thus be $N_{\rm bin} \sim 10^4-10^5$. However, $N_{\rm bin}$ is more likely to be $\lesssim 10$ in GCs \citep[e.g.][see also Appendix \ref{sec:app1}]{morscher15,kremer20} and $\sim 10^2-10^3$ in NCs \citep[e.g.][]{arcasedda20b}, thus suggesting that these environments are unlikely to be main contributor to the population of NS-BH mergers. Extending our calculations to YCs, instead, yields to:
\begin{eqnarray}
\displaystyle
    \Gamma_{\rm YC} & = &\left(\frac{\rho_{\rm MWEG}}{0.0116{\rm ~Mpc}^{-3}}\right)\left(\frac{N_c}{10^5}\right) N_{\rm bin}\times \\
    & & \times 
    \begin{cases}
    0.04 - 3.7 {\rm ~yr}^{-1} {\rm ~Gpc}^{-3}, & \sigma = 0.3{\rm ~km ~s^{-1}}, \nonumber \\
    0.1 - 12.3   {\rm ~yr}^{-1} {\rm ~Gpc}^{-3}, & \sigma = 1.0{\rm ~km ~s^{-1}}, \nonumber \\
    0.3 - 36.6   {\rm ~yr}^{-1} {\rm ~Gpc}^{-3}, & \sigma = 3.0{\rm ~km ~s^{-1}}, \nonumber \\
    \end{cases}
    \label{eq:YCrate}
\end{eqnarray}
with lower limits corresponding to $Z=0.02$ and the NSSTBH configuration. According to Eq. \ref{eq:2}, the $\sigma$ values adopted above correspond to a cluster mass $M_{\rm YC} = (280 - 3,166 - 28,000)\Ms$ assuming $r_h = 1$ pc. Note that such an optimistic rate is obtained assuming that all MW-like galaxies in the local Universe have a number of YCs $\sim 10^5$, each of which contains at least $N_{\rm bin} = 1$ binary that undergoes the type of scattering explored here.

\begin{table}
    \centering
    \begin{tabular}{cccc}
    \hline
        configuration & $Z$& $k$  & $\delta$\\        
        & &[Gyr$^{-1}$]&\\
    \hline
    \hline
        BHSTNS   & 0.0002 & $(5.2 \pm 0.9) \times 10^{-2}$ & $0.99 \pm 0.01$ \\
                 & 0.02   & $(3.6 \pm 0.2) \times 10^{-2}$ & $0.98 \pm 0.03$\\
        NSSTBH   & 0.0002 & $(1.0 \pm 0.2) \times 10^{-3}$ & $0.78 \pm 0.12$\\
                 & 0.02   & $(4.0 \pm 0.6) \times 10^{-4}$ & $0.89 \pm 0.08$ \\
    \hline
    \end{tabular}
    \caption{Best fitting parameters for clusters individual merger rate. Col. 1: scattering configuration. Col. 2: metallicity. Col. 3-4: parameters of the fitting function for the individual merger rate in Equation \ref{eq:1}.}
    \label{tab:tab1}
\end{table}

This rate falls within the LVC measurements \citep{190814} and is in remarkably good agreement with simulations of compact YCs \citep{rastello20}. Assuming instead similar number densities for YCs and GCs, $\rho_{\rm GC}=2.31$ Mpc$^{-3}$, leads to $\Gamma_{\rm YC}=(1.5\times 10^{-4} - 0.06)\yrgpc$, in agreement with recent results from \cite{fragione20} (for more details see Appendix \ref{sec:app3}).
The agreement between our models and $N$-body simulations, in spite of the different assumptions adopted, suggests that the dynamical formation of NS-BH binaries is driven mostly by stellar dynamics and is less affected by stellar evolution and post-Newtonian corrections. Our simplified approach enables us to produce a catalogue of $\sim 10^3$ NS-BH mergers, which can be used to constrain their overall properties, and to access the NCs mass range, for which full direct simulations are prohibitive.

\begin{table*}
    \centering
    \begin{tabular}{ccccc|cc|cc}
    \hline
                  & &  & & &\multicolumn{4}{c}{$\Gamma_{\rm ind}$ [Gyr$^{-1}$]}\\
                  \hline
                  & &  & & &\multicolumn{2}{c|}{NSSTBH} & \multicolumn{2}{c}{BHSSTNS}\\
    cluster & $M$  & $r_h$ & $\sigma$     & $N_{\rm bin}$ &\multicolumn{2}{c|}{$Z$} &  \multicolumn{2}{c}{$Z$} \\
       type & [$\Ms$] & [pc] &[km s$^{-1}$] &               &0.0002 & 0.02 & 0.0002 & 0.02 \\
    \hline
    \hline
YCs&$3\times 10^2$ &$1$ &$0.3^{1}$&$1$& $1.17\times 10^{-4}$ &$3.25\times 10^{-5}$ &$3.22\times 10^{-3}$ &$2.24\times 10^{-3}$\\
 
YCs&$3\times 10^3$ &$1$ &$1.0^{1}$&$1$& $2.99\times 10^{-4}$ &$9.57\times 10^{-5}$ &$1.06\times 10^{-2}$ &$7.37\times 10^{-3}$\\

YCs&$3\times 10^4$ &$1$ &$3.0^{1}$&$1$& $7.06\times 10^{-4}$ &$2.56\times 10^{-4}$ &$3.15\times 10^{-2}$ &$2.18\times 10^{-2}$\\
 
GCs&$10^5$         &$3$ &$5.5^{2}$&$1$& $1.13\times 10^{-3}$ &$4.41\times 10^{-4}$ &$5.75\times 10^{-2}$ &$3.96\times 10^{-2}$\\

NCs&$8\times10^6 $ &$3$ &$50^{3} $&$1$& $6.34\times 10^{-3}$ &$3.19\times 10^{-3}$ &$5.13\times 10^{-1}$ &$3.50\times 10^{-1}$\\

    \hline
    \end{tabular}
    \caption{Individual merger rate for different cluster types. Column 1-4: cluster type, mass, half-mass radius, and typical velocity dispersion. Col. 5: average number of binaries containing a NS or a BH. Col. 6-7: individual merger rate for NSSTBH configuration and different metallicities. Col. 8-9: individual merger rate for BHSTNS configuration and different metallicities. $^{1}$ \citep{piskunov07,soubiran18,jackson20}. $^{2}$\citep{harris10}. $^{3}$\citep{feldmeier14}.}
    \label{tab:tab2}
\end{table*}

\section{Dynamical formation of GW190814 and GW190426\_152155}
\label{sec:gw190814}

\begin{figure}
    \centering
    \includegraphics[width=\columnwidth]{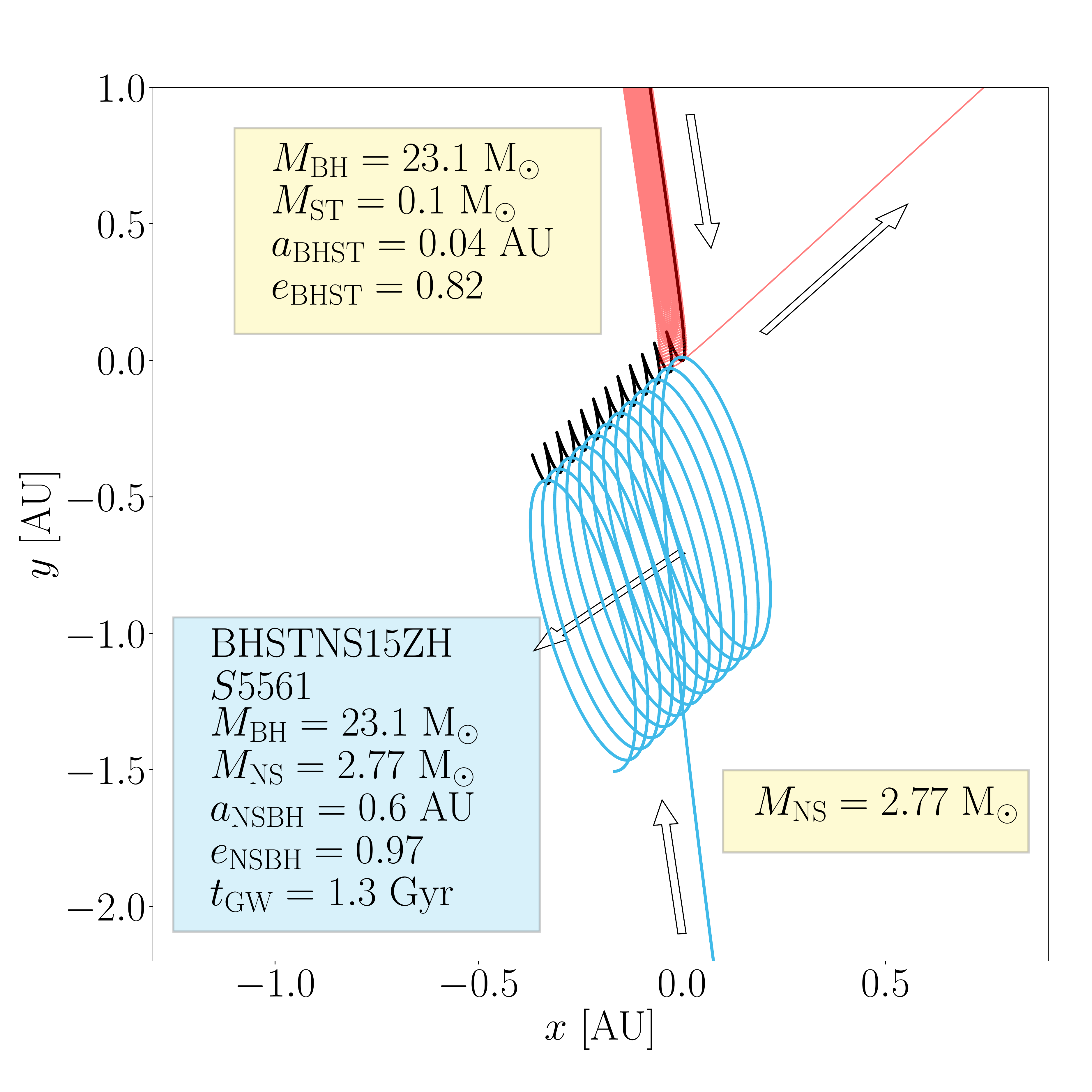}
    \caption{Formation of a GW190814 prototype in one of our models ($Z = 0.02$, $\sigma = 15$ km s$^{-1}$). During the scattering the NS swaps with the star (ST), which is ejected away, leading to the formation of a highly eccentric binary with merging time $t_\gw \lesssim 1.3$ Gyr.}
    \label{fig:GW190814}
\end{figure}

Among all our models we find 11 mergers with a BH with mass $20<M_\bh/\Ms<25$ and a NS with mass $M_\ns>2\Ms$. One interesting example is BHSTNS15ZH-S5561 (configuration BHSTNS, $\sigma = 15$ km s$^{-1}$, metallicity $Z = 0.02$), which has $M_\bh = 23.1\Ms$ and $M_\ns = 2.77\Ms$, i.e. within $\sim 0.4\%$ and $\sim 7\%$ from the GW190814 measured values. As shown in Figure \ref{fig:GW190814}, at formation BHSTNS15ZH-S5561 has a semimajor axis $a=0.6$ AU and eccentricity $e=0.97$, merging within $t_\gw = 1.3$ Gyr. Note that the large NS mass is due to the adopted mass spectrum for compact remnants, which enable the formation of NS with a maximum mass of $\sim 3\Ms$ at solar metallicity (see Appendix \ref{sec:app2}).

\begin{figure}
    \centering
    \includegraphics[width=\columnwidth]{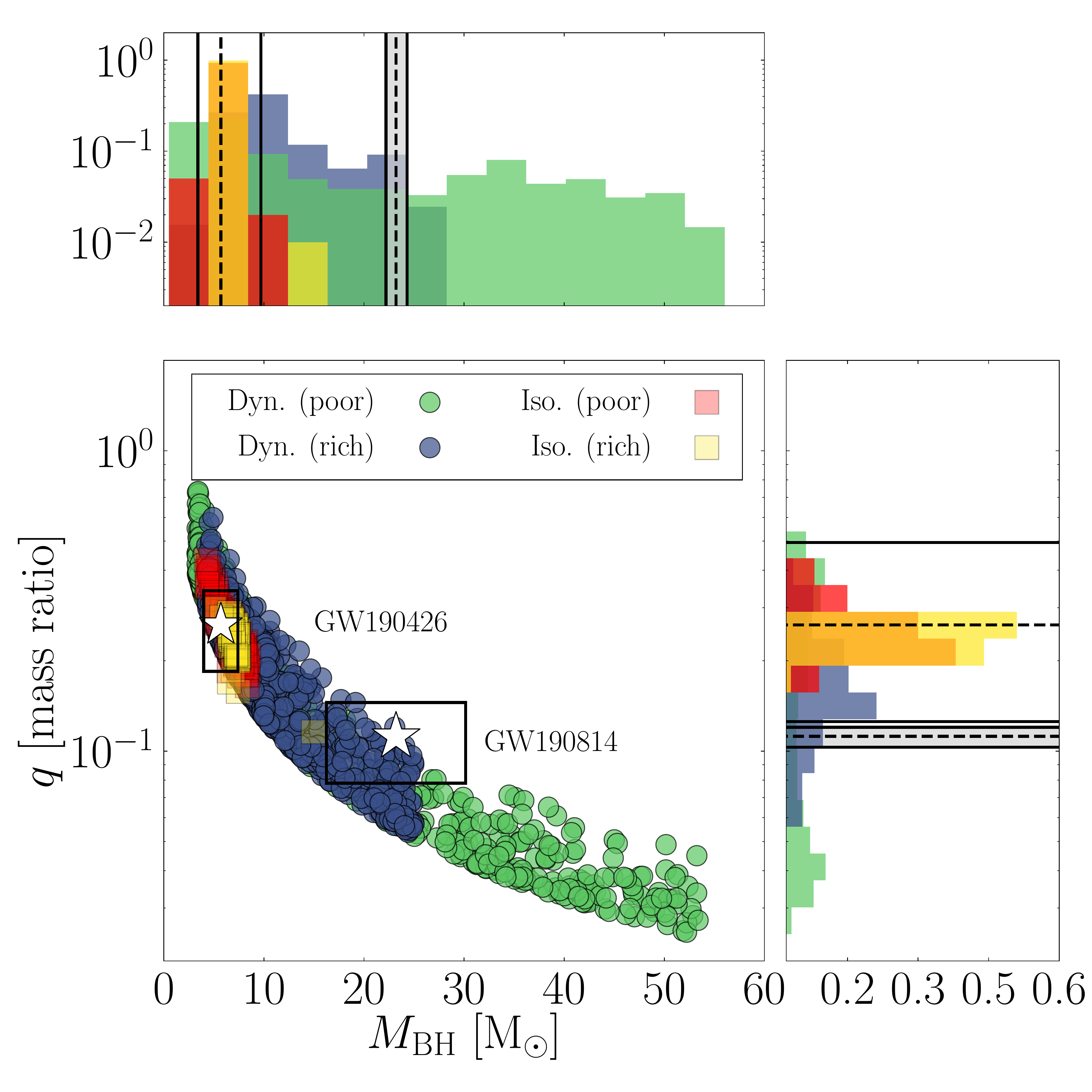}\\
    \caption{Mass ratio $q$ versus BH mass $M_{\rm BH}$ for 
    NS-BH mergers in our metal-poor (green dots) and metal-rich (purple dots) models, compared to the measured values for GW190814 and GW190426\_152155 (white stars). We include the combined distribution for metal-poor (red squares) and metal-rich (yellow squares) binaries derived from \cite{giacobbo18} (model $CC15\alpha5$). The black boxes enclosing GW190814 and GW190426\_152155 represent regions deviating $< 30\%$ from the observed $M_\bh$ and $q$. In the side histograms, the dotted lines identify the measured mass and mass ratio for both GW sources, while the shaded(empty) areas encompass the measured $90\%$ credible level for GW190814(GW190426\_152155).}
    \label{fig:qMbh}
\end{figure}

To identify other NS-BH mergers similar to GW190814, we use the BH mass $M_\bh$ and the mass ratio $q$. Figure \ref{fig:qMbh} compares these quantities for GW190814, our dynamical mergers, and isolated mergers \citep[adapted from][]{giacobbo18} for metal-poor ($Z = 0.0002$) and metal-rich ($Z=0.02$) stellar progenitors. It must be noted that \cite{giacobbo18} adopts a {\it rapid} SN explosion scheme \citep[see][]{fryer12}, whereas our model is based on a {\it delayed} SN scheme for the calculation of compact remnants' masses. Nonetheless, updated models of isolated binary evolution accounting for both rapid and delayed SN, which have shown a broad agreement with \cite{giacobbo18} models, suggest that the amount of mergers with properties similar to GW190814 is limited to $<0.1-1\%$, regardless of the SN mechanism considered \citep{zevin20}. 

To identify systems similar to GW190814 in our database, we shortlist all mergers having a mass and mass ratio within the $30\%$ of the measured values for GW190814. We find a probability $P_\lvk =4.1 - 5.9\%$ to find GW190814-like mergers in metal-poor configurations BHSTNS and NSSTBH, respectively, and $P_\lvk = 8.2 - 11.3\%$ for solar metallicity models. From Equation \ref{eq:YCrate} and Table \ref{tab:tab2} we can thus derive a rate for mergers similar to GW190814 in YCs as:
\begin{eqnarray}
\displaystyle
\Gamma_\lvk & = & P_\lvk \Gamma_{\rm YC} = \\
    & =  &
    \begin{cases}
            ( 0.01  - 2.1 ) N_{\rm bin} \yrgpc  & Z / {\rm Z}_\odot= 0.01, \nonumber \\
            ( 0.009 - 2.9 ) N_{\rm bin} \yrgpc  & Z / {\rm Z}_\odot= 1, \nonumber 
    \end{cases}
\end{eqnarray}
with the lower(upper) limits corresponding to the case $\sigma = 1$($3$) km s$^{-1}$, i.e. the typical value of velocity dispersion for MW young and open clusters\citep{soubiran18,kuhn19,jackson20}, and to configuration NSSTBH(BHSTNS). 
 
Using the same procedure, we also seek for mergers similar to GW190426\_152155, a NS-BH merger detected during the O3 LVC observation run, characterised by $M_\bh = 5.7^{+4.0}_{-2.3}\Ms$ and $M_\ns = 1.5^{+0.8}_{-0.5}\Ms$. We find mergers with primary mass and mass ratio within $30\%$ of the measured value for GW190426\_152155 in $15\%$ of our models regardless the progenitor metallicity, thus indicating that dynamical mergers can produce a substantial fraction of systems with a relatively low mass. Note that the interval of $M_\bh$ and $q$ values assumed in this case falls well within the observed $90\%$ credible interval level.

\begin{figure*}[!h]
    \centering
    \includegraphics[width=0.9\columnwidth]{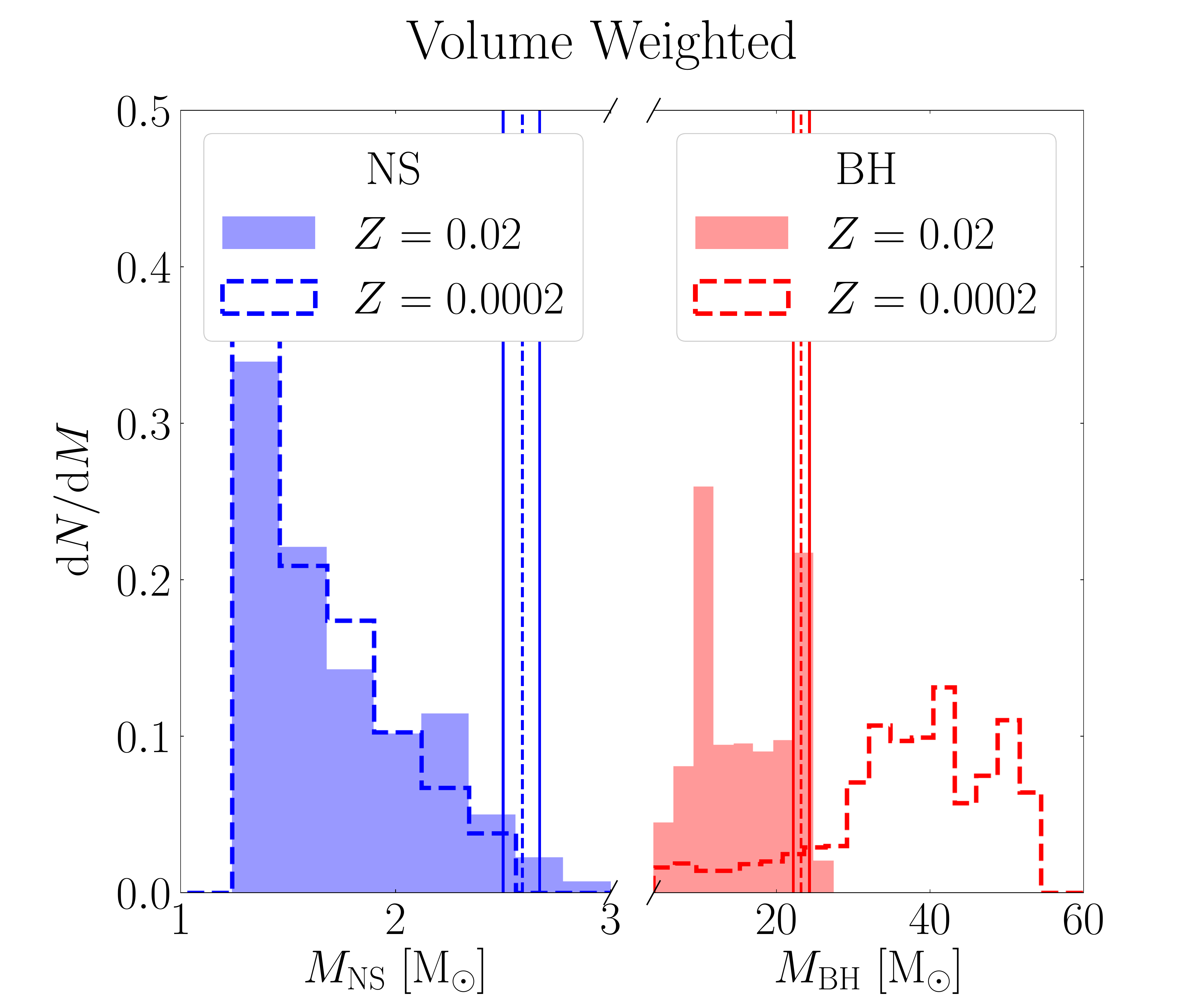}\\
    \includegraphics[width=0.9\columnwidth]{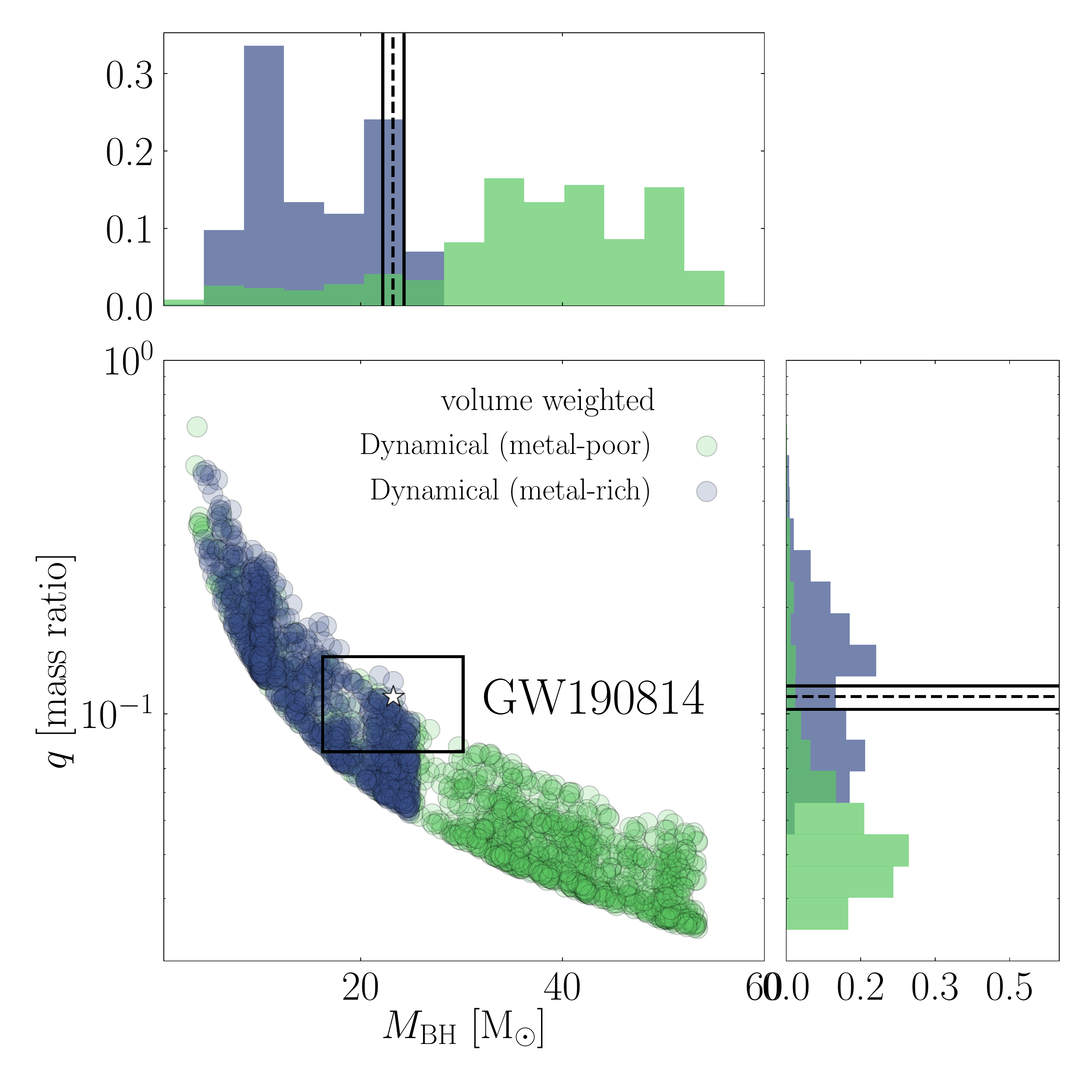}\\
    \includegraphics[width=0.9\columnwidth]{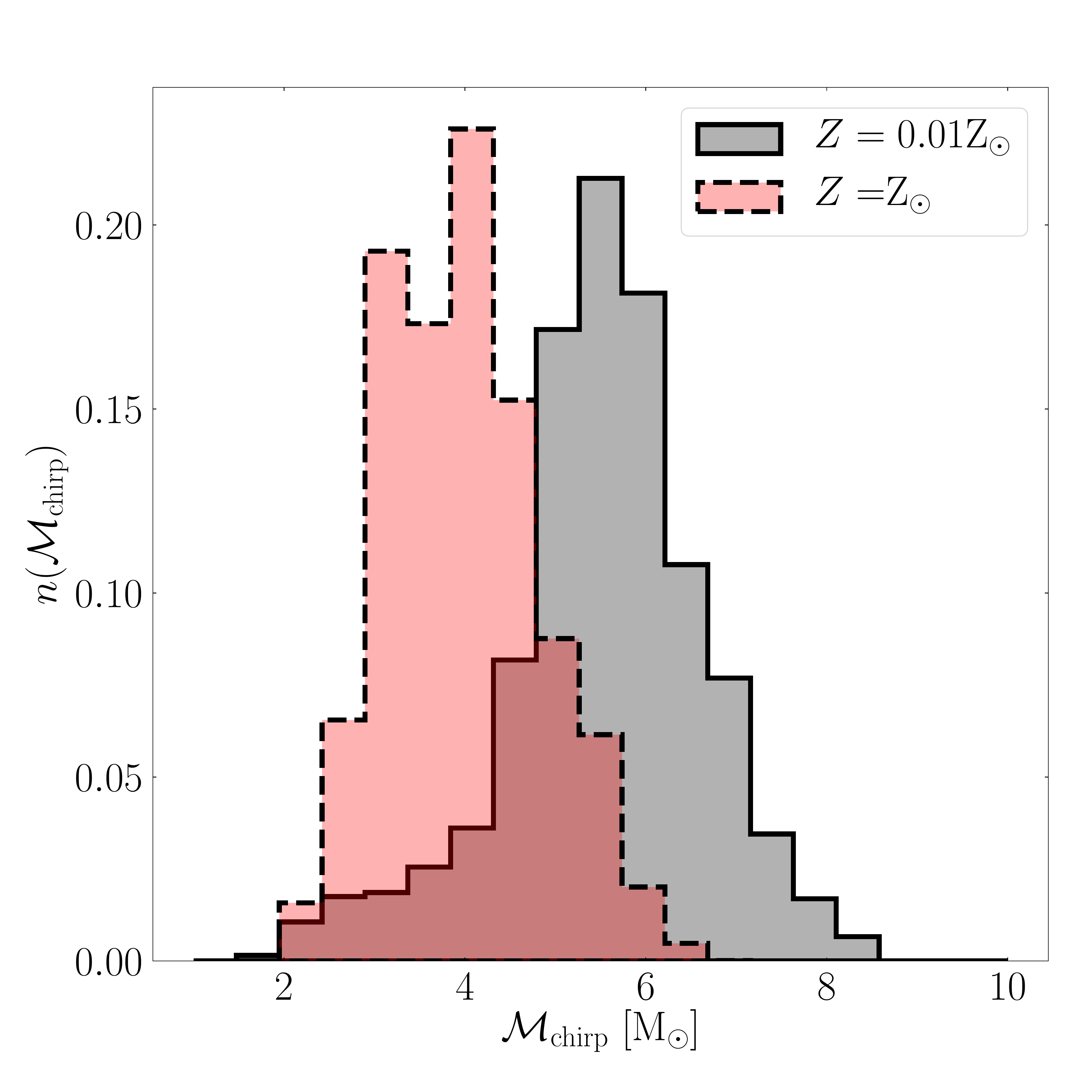}\\
    \caption{Top: volume weighted mass spectrum for merging NSs (left, blue) and BHs (right, red) for a metallicity $Z=0.0002$ (open dashed steps) and $Z=0.02$ (filled steps). The vertical lines mark the value measured for GW190814 and corresponding uncertainties. Central: as in Figure \ref{fig:qMbh}, but here the mass ratio and primary mass distribution are weighted with two selection functions to mimic the dependencies affecting the detector accessible volume. Bottom: mergers chirp mass for metal-rich (filled red steps with dashed edge) and metal-poor (filled grey steps with straight edge) clusters. }
    \label{fig:qMbhV}
\end{figure*}

The analysis above does not account for potential observation biases that can affect GW detectors. For instance, the volume within which LIGO can detect a given class of sources depends on several parameters, like the source mass and mass ratio, the distance, the sky location, or the mutual inclination of the spins. For binaries with a total mass $M_1+M_2 = (10-100)\Ms$, \cite{fishbach17} showed that this volume scales with the primary mass following a power-law $VT\propto M_1^{2.2}$ and decreases at decreasing the mass ratio $q$. Using Figure 1 in \cite{fishbach17}, we extract $VT$ and $q$ at a fixed primary mass value $M_1 =(10,~20,~25,~30,~50)\Ms $, finding that such relation is well described by a power-law $VT\propto q^{\beta}$, with $\beta \sim 0.4-0.6$. In the following, we adopt a slope $\beta = 0.5$, which is the value associated with a primary mass $M_1\sim 20\Ms$. As we show in Appendix \ref{sec:app4}, setting $q = 0.4$ or $q=0.6$, i.e. the values typical of systems with a primary $M_1<35\Ms$, leads our estimated merger rate to vary by less than $10\%$. To mimic the selection effect connected with the binary primary and mass ratio, we augment our population of NS-BH mergers to 10,000 by sampling them from the combined $M_\bh-M_\ns$ distribution and, from the augmented sample, we extract 1,000 NS-BH mergers weighing the probability to select a given mass and mass ratio with the selection functions  $f_M = k_\bh M_\bh^{2.2}$ (with $k_\bh$ a normalization constant), and $f_q = k_q q^{0.5}$ \cite[see also][]{arcasedda19b, arcasedda20b}.
The resulting \textit{volume weighted} mass distribution for BHs and NSs, and the $M_\bh-q$ plane are shown in Figure \ref{fig:qMbhV}. The percentage of mergers falling inside the limiting values adopted for GW190814 remains limited to $P_\lvk =  4.3 \pm 0.4\%$ for metal-poor clusters, owing to the fact that heavier BHs have a larger probability to be selected. For metal-rich environments and the BHSTNS configuration, instead, this probability increases to $P_\lvk = 22 \pm 2 \%$, leading to an optimistic ``volume-weighted'' merger rate for GW190814-like mergers in YCs of 
\begin{eqnarray}
\displaystyle
    \Gamma_{\rm\gws,V} & = &\left(\frac{\rho_{\rm MWEG}}{0.0116{\rm ~Mpc}^{-3}}\right)\left(\frac{N_c}{10^5}\right) N_{\rm bin}\times \\
    & & \times 
    \begin{cases}\nonumber
    (0.008 - 0.6)  \yrgpc & \sigma = 0.3{\rm ~km ~s^{-1}},\\
    (0.023 - 2.0)  \yrgpc & \sigma = 1.0{\rm ~km ~s^{-1}},\\
    (0.062 - 5.8)  \yrgpc & \sigma = 3.0{\rm ~km ~s^{-1}},\\
    \end{cases}
\end{eqnarray}
at redshift $z<1$ and adopting $\rho_{\rm MWEG} = 0.0116$ Mpc$^{-3}$, $N_c = 10^5$, and $N_{\rm bin}= 1$ as scaling values. The lower(upper) limit corresponds to configuration NSSTBH(BHSTNS). The same calculation for metal-poor clusters yields to a maximum value of $0.005-1.7\yrgpc$, thus suggesting that a population of metal-rich YCs with $\sigma = 1-3$ km s$^{-1}$, i.e. $r_h \sim 0.5-1.5$ pc and $M \sim 10^3-10^4\Ms$, represent the most suited class of environments to explain the origin of GW190814. In comparison, the isolated scenario predicts a merger rate of $<0.1\yrgpc$, regardless of the assumptions on the SN mechanism \citep{zevin20}.

In the case of GW190426\_152155, our procedure leads to $P_\lvk = 1.8 - 4.7\%$, with the lower(upper) limit referring to $Z=0.02$($0.0002$), thus indicating that the contribution of dynamical mergers to the population of low-mass NS-BH mergers can be non negligible. 

The merger rates above represent optimistic estimates that rely on the assumption that YCs: 1) have around solar metallicity, 2) have all the same velocity dispersion, and 3) are $\sim 10^5$ in MW-like galaxies. The discovery of other mergers similar to GW190814, i.e. with chirp masses $\gtrsim 4\Ms$ and mass ratio $<0.1$ could help in placing constraints on the processes that regulate the formation and evolution of young clusters in the local Universe. 

Using the volume weighted catalogue, we calculate the percentage of mergers with a primary mass in the ranges $[<~7, ~7-15, ~\geq~15]\Ms$, and a companion mass in the range $[<~2, ~2-2.5,~ >~2.5]\Ms$. As summarized in Table \ref{tab:tab3}, we find that a dominant contribution to NS-BH mergers from metal-rich clusters would result in a $96\%$ probability to detect a primary heavier than $>15\Ms$, whereas for metal poor clusters this probability is comparable for {\it light} and {\it heavy} primary components.

\begin{table*}
    \centering
    \begin{tabular}{c|ccc|ccc|c}
    \hline
	$Z$ & \multicolumn{3}{c|}{$M_\ns$ [$\Ms$]}& \multicolumn{3}{c|}{$M_\bh$ [$\Ms$]}& $(M_\bh ,~M_\ns)$ [$\Ms$] \\
		& $< 2$ & $~2-2.5$ & $~\geq 2.5$ & $< 7$ & $~7-15$ & $~\geq 15$ & $>7 $ and $>2$\\  
	\hline	
	\hline
	0.0002 & $77.8\%$ & $19.5\%$& $2.7\%$ & $4.0\%$& $40.9\%$& $55.2\%$ &$13.5\%$\\
	0.02   & $87.4\%$ & $12.3\%$& $0.3\%$ & $1.3\%$& $ 3.3\%$& $95.5\%$ &$23.9\%$\\
	\hline
    \end{tabular}
    \caption{Occurrence of NS-BH mergers in different mass ranges. Col. 1: metallicity. Col. 2-4: probability to have a NS mass within a given mass range. Col. 5-7: the same as for columns 2-4, but for BHs. Col. 8: probability to have a merger with a primary mass $>7\Ms$ and a companion mass $>2\Ms$.}
    \label{tab:tab3}
\end{table*}

In the extreme case in which NS-BH mergers form dynamically and mostly in metal-rich clusters, the estimate above implies that 9 out of 10 detections of NS-BH mergers would involve a BH with $M_\bh > 15\Ms$, and 2-3 among them will have a companion with mass $>2\Ms$. Comparing these predictions with expectations from other channels and actual detections can help in unravelling the markers of different formation channels in detected sources and shed a light on the role of dynamics in determining the assembly of NS-BH mergers.

\section{Summary and conclusions}

In this Letter we exploited a suite of 240,000 $N$-body simulations of hyperbolic encounters in star clusters to investigate the dynamical formation of NS-BH mergers with properties similar to GW190814 and GW190426\_152155. Our main results can be summarized as follows:
\begin{enumerate}
\item We find that the NS-BH merger probability depends strongly on the star cluster velocity dispersion, following a power law with slope $\delta \sim 0.6-1.1$. Overall, around $0.5\%$ of our models lead to a NS-BH merger;

\item We derive an individual merger rate, i.e. number of mergers per time unit, for typical NCs (up to $\sim 0.5$ mergers per Gyr), GCs ($<0.06$ Gyr$^{-1}$), and YCs $(<0.01)$ Gyr$^{-1}$;

\item Since YCs outnumber GCs and NCs by a factor up to $10^5$ in typical galaxies, they might be the major contributor to the population of dynamical NS-BH mergers. In the local Universe, we infer a NS-BH merger rate for YCs of $\Gamma_{\rm YC} = (0.04-36)\yrgpc$;

\item Among all simulations, we identify $\sim 5-10\%$ NS-BH mergers with a BH mass and mass ratio compatible with GW190814, and $\sim 15\%$ with properties similar to GW190426\_152155. We exploit our models to derive a ``raw'' merger rate for dynamically formed GW190814-like sources, i.e. with mass and mass-ratio within $30\%$ the observed values, of $\Gamma \sim (0.01-2.9) \yrgpc$;

\item To place our models in the context of LVC detections, we assume that the probability to select a NS-BH merger in our sample depends on the primary mass ($\propto M_{\rm BH}^{2.2}$) and mass ratio ($\propto q^{0.5}$) to mimic the potential selection effects to which GW detectors might be subjected. This ``volume weighted'' sample contains only $\sim 4.3 \pm 0.4\%$ of GW190814-like systems in metal-poor clusters, but this percentage becomes noticeable in metal-rich clusters ($\sim 22 \pm 2\%$);

\item Combining the volume weighted sample of mergers with the large abundance of YCs in Milky Way-like galaxies, we derive an optimistic rate for GW190814-like mergers of $0.008-5.8\yrgpc$ at low redshift, in the ballpark of LVC predictions and up to 100 times larger than the estimates obtained from isolated binary stellar evolution models;

\item Using the same selection procedure, we find $\sim 1.8-4.7\%$ of mergers with properties comparable to GW190426\_152155, with the lower limit corresponding to metal-poor clusters. This relatively low occurrence disfavours, but does not rule out, a dynamical origin for GW190426\_152155;

\item We suggest that the mass spectra of compact remnants in NS-BH merger candidates can be used to identify markers of different formation channels. In the extreme case in which all mergers formed dynamically in metal-rich clusters, we predict that 9 out of 10 mergers should involve a BH with $M_\bh > 15\Ms$ and at least 2 of them involve a companion with mass $>2\Ms$. Comparing these predictions with future detections can shed a light on the dynamical channel and provide new insights on the mass spectrum of compact objects in the lower mass-gap range. 
\end{enumerate}

\section*{Acknowledgements}
MAS acknowledges financial support from the Alexander von Humboldt Foundation for the research program ``The evolution of black holes from stellar to galactic scales'', the Volkswagen Foundation Trilateral Partnership through project No. I/97778 ``Dynamical Mechanisms of Accretion in Galactic Nuclei'', and the Deutsche Forschungsgemeinschaft (DFG, German Research Foundation) -- Project-ID 138713538 -- SFB 881 ``The Milky Way System''. MAS is grateful to Martina Donnari for providing useful comments to an earlier version of this manuscript.

\appendix

\section{Initial conditions I. Binary-single interaction rates}
\label{sec:app1}

The idea at the basis of our approach is that BH-NS binaries form via interaction of a free roaming single compact object (a BH or NS) and another compact object (a NS or BH) paired with a star. Here, we consider ``NSs'' all compact objects with a mass $<3\Ms$ and ``BHs'' otherwise. Since the two compact objects are heavier than the star, on average this configuration favours the ejection of the least massive component and the formation of a binary with a higher binding energy \citep{sigurdsson93}. 
We explore two different configurations: NSSTBH (NS-star binary impacting over a single BH), and BHSTNS (BH-star binary impacting over a single NS), and we vary the stellar metallicity to either $Z=0.0002$ or $Z=0.02$ and the cluster velocity dispersion to $\sigma = 5-15-20-35-50-100$ km s$^{-1}$, thus covering the range of values going from young massive clusters (YCs), to globular clusters (GCs) and nuclear clusters (NCs). 

In such stellar ensembles, the interaction rate between a binary with component mass $M_{1,2}$, semimajor axis $a$, and eccentricity $e$, and a single object with mass $M_3$ can be written as 
\begin{equation}
\der  R / \der t = \dot{R} = N_{\rm bin}n\sigma\Sigma,
\label{eq:rate}
\end{equation}
being $N_{\rm bin}$ the average number of binaries coexisting in the cluster that contain either an NS or a BH, $n$ the density of scattering stars, $\sigma$ the environment velocity dispersion, and $\Sigma$ the binary cross section
\begin{equation}
\Sigma = \mathrm{\pi} a^2 \left(1-e \right)^2 \left[ 1+\frac{2G(M_1+M_2+M_3)}{\sigma^2a(1-e)} \right].
\end{equation}

The density of a population of $N_{\rm CO}$ (either BHs or NSs) with mean mass $\langle M_{\rm CO} \rangle$ inhabiting a cluster with mass $M$ and half-mass radius $R_h$ can be estimated as $n_{\rm CO} \sim N_{\rm CO} / (R_\mathrm{CO}^3)$. For BHs, this quantity can be inferred for instance from recent studies on BH retention fraction and consequent formation of a tight BH subsystem \citep{breen13,morscher15,arcasedda18a}. In this work, we assume that the segregated population of BHs has a density comparable to the overall density of the cluster, $n_{\rm BH} \sim M/(\langle M_{\rm BH} \rangle R_h^3)$, as suggested in \cite{arcasedda18a}.
For NSs instead, we consider the fact that mass-segregation is prevented by the presence of BHs in the cluster centre, and that their total mass is around 0.01 times the cluster mass. Thus, we adopt $n_{\rm NS} = 0.01 n_{\rm GC}$ as an upper limit on the average density of NSs. 
The number of binaries in the cluster is a crucial parameter. 
To bracket this quantity, we take advantage of the MOCCA Survey Database I \citep{askar17}, a suite of around 2,000 Monte Carlo models of star clusters with initial masses in the range $M = 2.4\times 10^4 - 7.7\times 10^5 \Ms$, thus covering the mass range of massive YCs and GCs. 

\begin{figure}
    \centering
    \includegraphics[width=0.75\columnwidth]{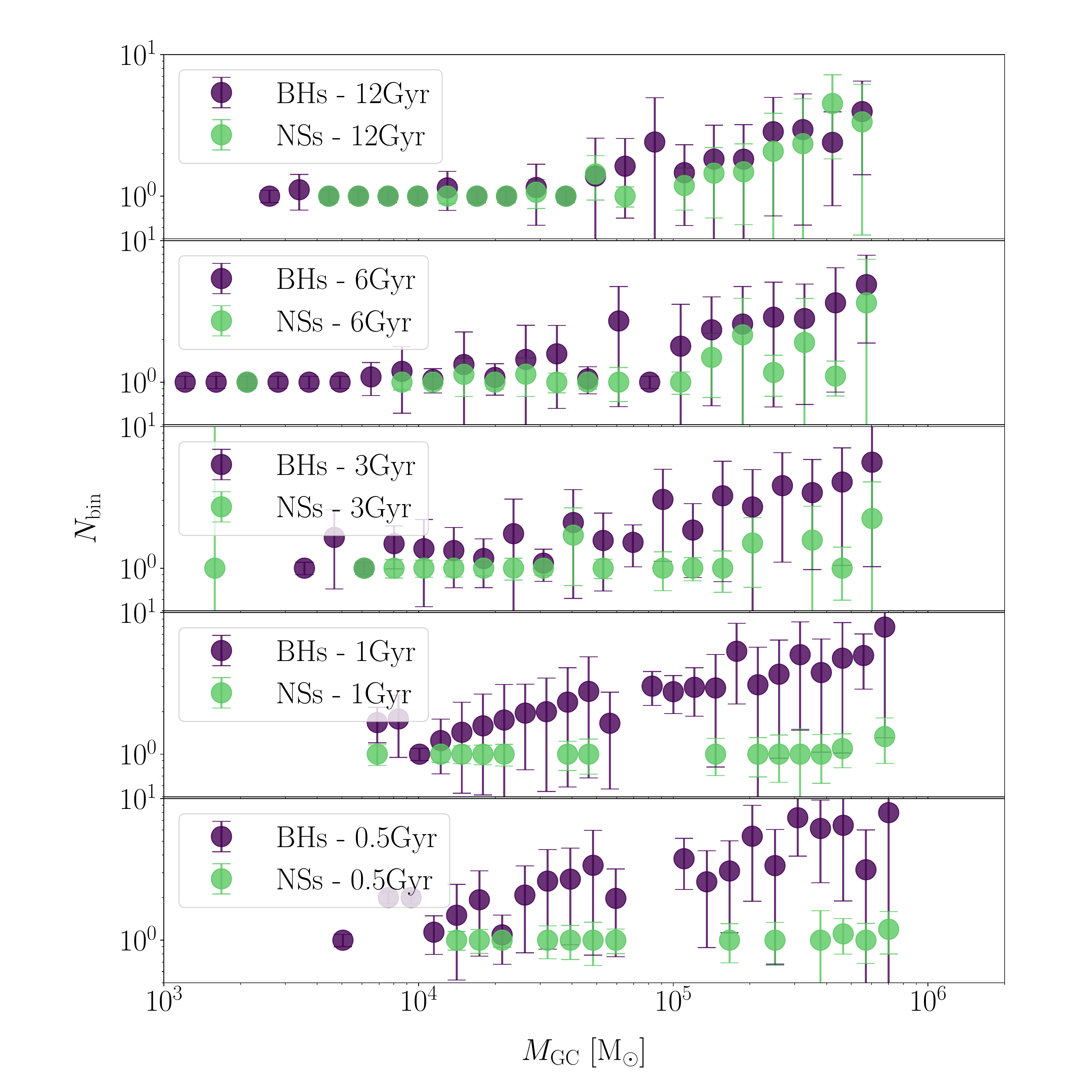}
    \caption{Average number of BH and NS in binaries as a function of the cluster mass and for different times, from top to bottom the time considered is $12~-~6~-~3~-~1~-~0.5$ Gyr, respectively. }
    \label{fig:appB}
\end{figure}

Using the MOCCA models, we calculate the average number of binaries containing a BH(NS) in cluster at different times ($0.5-1-3-6-12$ Gyr respectively) and in different mass bins, as shown in Figure \ref{fig:appB}. 

The figure highlights a clear, although non-trivial, dependence between $N_{\rm bin}$ and the cluster mass.
Clusters lighter than $M < 10^5\Ms$ are characterised by $N_{\rm bin}\sim 1$ for NS in binaries regardless of the cluster mass, while it ranges in between $N_{\rm bin}\sim 1-3$ for BHs, especially at earlier times. 
This is likely due to the fact that low-mass clusters have shorter relaxation times, thus the formation of binaries containing BHs and their ejection via strong scatterings occur earlier than in heavier mass clusters. At values $M>10^5\Ms$ instead, the number of BH in binaries varies between $N_{\rm bin} = 1-10$, almost regardless of the time, whereas the number of NS in binaries tends to be smaller $N_{\rm bin} = 1-4$, especially at earlier times. Given this non-trivial behaviour, in our calculations we leave $N_{\rm bin}$ as a scaling value. 

We note that this assumption is compatible with results of other models \citep[see e.g.][]{morscher15}, even the most recent one that implements updated stellar evolution for binary and single stars, new prescriptions for SN explosion mechanisms and recoil kicks, and post-Newtonian formalism for compact object interactions \citep[e.g.][]{kremer20}, thus suggesting that the number of binaries left in the cluster at late evolutionary stages is likely the result of dynamics, rather than other mechanisms that, on the other hand, can affect the cluster structure. 

Under the set of assumptions above, adopting a cluster velocity dispersion of $\sigma = 5$ km s$^{-1}$, and using the median values of the masses of the binary and third object and the binary semimajor axis and eccentricity, we estimate an interaction rate of $\dot{R} \sim 2-4$ Gyr$^{-1}$ for configuration NSSTBH and  $\dot{R} \sim 150-400$ Gyr$^{-1}$ for configuration BHSTNS. 

To check the reliability of our calculations, we compare our results with MOCCA models as described in our companion paper \citep{arcasedda20}, finding a range of values fully compatible with our theoretical estimate.

\section{Initial conditions II. Mass and orbital properties of binary-single scattering experiments}
\label{sec:app2}

The binary-single scattering configuration explored in this work is characterised by the masses of the three objects (star, BH, and NS), the orbit of the initial binary, and the orbital properties of the single-binary interaction.

We sample the zero age main sequence (ZAMS) mass of the three objects from a Salpeter mass function. At metallicity values $Z = 0.0002$($0.02$), we assume that all stars with a ZAMS mass $M_{\rm min BH} > 20.5$($18$)$\Ms$ evolve into BHs \citep{belczynski02,spera17}, whereas lighter stars with a mass above $M_{\rm min NS} > 8$($6.5$)$\Ms$ evolve into NSs. 

For stars, we select only masses in the range $0.1-1$ M$_\odot$. This choice is motivated by the fact that the timescale associated with the binary single scatterings modelled here is 10-100 times longer than the half-mass relaxation time of the host cluster (see Appendix \ref{appC}), thus generally longer than the evolutionary time for stars with a mass $m>2$ M$_\odot$, which is $t_{\rm age} = 10{\rm ~Gyr} (m/{\rm M}_\odot)^{-2.5} \lesssim 1-1.5$ Gyr, depending on the metallicity. 
Stars with a mass $1<M_{\rm st}/\Ms<2$, which constitute $2\%$ of the whole stellar population, evolve on timescales ($1.5-9$ Gyr) comparable to the cluster relaxation time, depending on the cluster velocity dispersion. For these stars, a proper modelling should also consider the evolutionary stage of the star, a feature that cannot be accounted for in our $N$-body models. Given the fact that they constitute a little fraction of the stellar population, as stars with a mass $<1\Ms$ constitute over the $95.5\%$ of the whole population, and that their stellar evolution timescale is close (or even longer) than the typical time for the scatterings studied here, we exclude $1-2\Ms$ stars from our models.

\begin{figure}
    \centering
    \includegraphics[width=0.4\columnwidth]{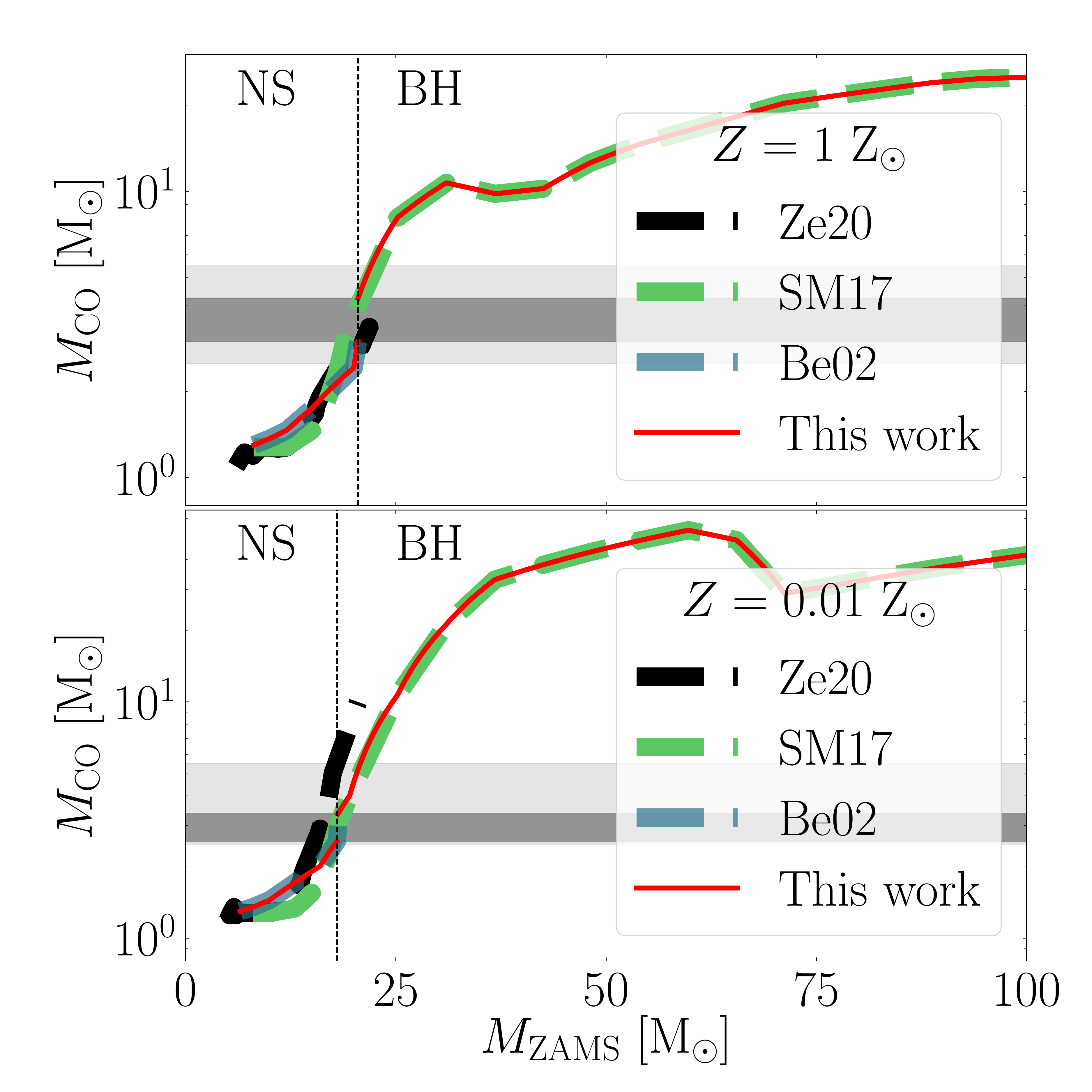}
    \includegraphics[width=0.4\columnwidth]{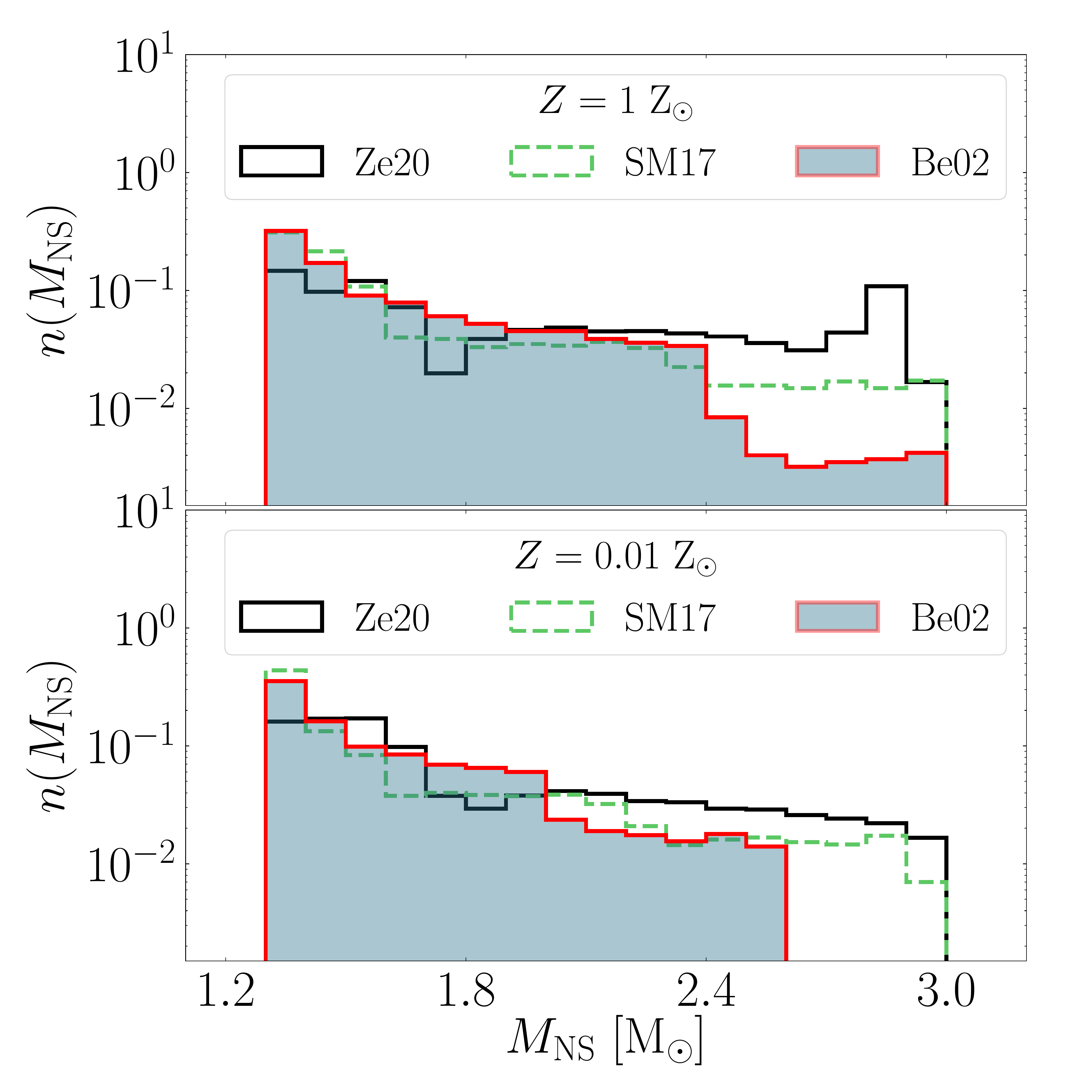}
    \caption{Left panel: final mass for compact objects as a function of the ZAMS mass adopted in this work (straight thin red line) in comparison with Ze20 (dashed thick black line), SM17 (dashed thick light green line), and Be02 (dashed thick light blue line) models. The light grey area encompass the supposed lower mass-gap, whereas the dark grey area labels the mass-gap featured in our models. Right panel: NS mass spectrum adopted in this work (thick red line) compared with Ze20 (black steps), SM17 (dashed green steps), and Be02 (filled blue steps).}
        \label{fig:imf}
\end{figure}

For BHs, we adopt the mass spectrum described in \cite{spera17} (hereafter SM17). In SM17 the authors use the SEVN tool. This tool implements prescriptions for single stellar evolution that include several SN explosion mechanisms and a treatment for pair instability and pulsational pair instability SNe, which naturally lead to a dearth of compact remnants with a mass in the range $65-120\Ms$ (the so-called upper mass gap). To derive the BH mass from the ZAMS mass we exploit Table 1-3 in \cite{spera17}\footnote{The SEVN code was note publicly available when this study began.}, according to which the compact remnant mass is calculated adopting the delayed SN explosion mechanism \citep{fryer12}. 

For NS, instead, we adopt the single star stellar evolution model from \cite{belczynski02} (hereafter Be02) implemented in the BSE package \citep{hurley02}. The maximum NS mass in our models reaches $\sim 2.5$($3$) M$_\odot$ for $Z=0.0002$($0.02$), with a small fraction of NSs ($0.03-0.05$) at solar metallicities having a mass $M_\ns > 2.3\Ms$, thus enabling us to explore the lower mass-gap region.

The combined use of SM17 prescriptions for BHs and Be02 for NSs leads our simulations to naturally exhibit a narrow lower mass-gap in the range $(3-4.2)\Ms$ at solar metallicity and $(2.6-3.1)\Ms$ for metal-poor systems. 
Figure \ref{fig:imf} compares the relation between the ZAMS and the remnant masses and the compact remnants' mass spectrum adopted here, in SM17, in Be02, and the new single stellar evolution models described in \cite{zevin20} (hereafter Ze20). We note that the mass spectrum adopted here is half-way between the case in which a wide lower mass-gap do exist and that in which NSs and BHs are linked by a continuous mass spectrum, and can provide a description of how the delayed SN mechanism or, more in general, an explosion mechanism that leads to a narrower mass-gap, can impact the properties of dynamically formed NS-BH mergers. In fact, our simulations suggest that an excess of $3-5\%$ of compact remnants with masses in the range $2.3-3$ M$_\odot$ can lead to a dynamical merger rate compatible with LIGO expectations. 

We note that the delayed SN model adopted here is only one among many possibilities, e.g. rapid SN mechanism \citep{fryer12}, electron-capture SN \citep{podsiadlowski04}. Nonetheless, none of the current models in the literature is capable of capturing the complex phases of SN physics, especially in the case of core-collapse SN, which can be altered significantly by stellar rotation \citep{mapelli19} and requires full 3D hydrodynamical simulations to be fully unveiled \citep[e.g.][]{burrows19}. From the theoretical point of view, recent single and binary stellar evolution population synthesis suggest that matching the GW190814 features requires that SN explosion proceeds on timescale longer than typically assumed \citep{zevin20}.
Moreover, observations of NSs and BHs detected through their electromagnetic counterparts seem to be inconclusive about the existence of a lower mass-gap, suggesting that it might be populated of both massive NSs \citep{Freire08} and light BHs \citep{giesers18,Thompson19}. This is also suggested by recent measurements based on microlensing events detected by the OGLE survey coupled with GAIA DR2 data, which favor a continuous mass spectrum in the $2-5\Ms$ mass range, rather than a lower mass-gap \citep{Wyrzykowski20}.

Regarding the orbital properties of the binary and the incoming object, as detailed in our companion paper \citep{arcasedda20} we
assume that the binary-single interaction is hyperbolic and in the regime of strong deflection. For the binary, we adopt as minimum semimajor axis the maximum between 100 times the star’s Roche lobe and 1000 times the ISCO of the compact object in the binary, to avoid that the star is disrupted or swallowed before the scattering takes place. The maximum semimajor axis allowed is, instead, calculated as the minimum between the hard-binary separation \citep{heggie75} and the relation suggested by \cite{rodriguez16}, who have shown that dynamically processed binaries have typical semimajor axis proportional to the binary reduced mass $\mu$ and the ratio between the cluster mass and semimajor axis, namely $a \sim k_d\mu M / R_h$. We adopt $k_d = 10$, which produces semimajor axis distribution in full agreement with binary-single scatterings of this kind found in MOCCA simulations with a $\sigma = 5$ km s$^{-1}$ \citep[see Figure 10 in][]{arcasedda20}.

\section{Criteria for the identification of merger candidates.}
\label{appC}
To identify NS-BH merger candidates we refine the selection procedure described in \cite{arcasedda20} as follows. We first calculate the GW timescale $t_\gw$ for all binaries, assuming 
\citep{peters64}
\begin{equation}
t_\gw =  \frac{5}{256}\frac{c^5 a_f^4 f(e_f)}{G^3M_1M_2(M_1+M_2)},
\label{peters}
\end{equation}
and
\begin{equation}
f(e) = \frac{(1-e^2)^{7/2}}{1+(73/24)e^2+(37/96)e^4}.
\end{equation}

We mark all NS-BH binaries with $t_\gw<14$ Gyr as "merger candidates". 
To determine whether these candidates can undergo merger in a cluster environment we need to infer the time at which the scattering takes place, i.e. the NS-BH binary formation time $t_f$, and the timescale over which the binary can get disrupted, e.g. via further strong encounters or secular perturbations. 

In a real cluster, the NS-BH binary formation time $t_f$ depends on a number of factors: the mass segregation time-scale, the core-collapse process, the formation of binaries and multiples in the cluster core, the formation or not of a BH subsystem. All these features are not captured by our three-body models, but are naturally accounted for in the MOCCA models. Therefore, we use a multi-stepped approach exploiting these high-resolution Monte Carlo models. First, we use the MOCCA database to reconstruct the logarithmic distribution of the ratio between $t_f$ and the half-mass cluster relaxation time calculated at 12 Gyr for all NS-BH binaries formed in MOCCA models, 
\begin{equation}
    {\rm Log} \tau \equiv {\rm Log} (t_f/ t_r),
    \label{tau}
\end{equation}
where the relaxation time is calculated as \citep{bt}
\begin{equation}
t_r =  \frac{0.65{\rm ~Gyr}}{{\rm log} (\Lambda)} \sqrt{\frac{M_c }{ 10^5\Ms}} \left(\frac{r_h}{1 {\rm ~pc}}\right)^{3/2}  .
\label{trelax}
\end{equation}
This enables us to provide an estimate of $t_f$ for a given value of the cluster relaxation time, which is directly connected with the cluster velocity dispersion, mass, and half-mass radius \citep{bt}.

The disruption of the NS-BH binary can be driven by either {\it impulsive} mechanisms, e.g. due to a strong encounter with another compact object, or {\it diffusive} mechanisms, e.g. due to the effect of the continuous interactions with passing-by stars or the mean field of the cluster. 

If the binary is soft, i.e. $G(M_1+M_2)/(2\sigma^2a) > 1$ \citep{heggie75}, we can distinguish between a {\it catastrophic} regime, i.e. the binary is disrupted in a single interaction, and a {\it diffusive} regime, i.e. the binary is disrupted owing to the secular effect impinged by interactions with cluster stars. A catastrophic interaction occurs if the impact parameter falls below a maximum value \citep{bt}
\begin{equation}
    b_{\rm max} = 1.5 a \left(\frac{GM_p}{(M_1+M_2)\sigma^2a}\right)^{1/4},
    \label{bmax}
\end{equation}
otherwise, the binary evolution is dominated by the secular, diffusive, mechanism.

In the catastrophic regime, the binary disruption occurs over a timescale \citep{bahcall85}
\begin{equation}
t_{\rm cat} = 4.5{\rm ~Gyr}\left(\frac{k_{\rm cat}}{0.07}\right)\left(\frac{3000 \Ms{\rm pc}^-3}{\rho}\right) \left(\frac{M_1+M_2}{20\Ms}\right)^{1/2}\left(\frac{10{\rm~AU}}{a}\right)^{3/2}. 
\end{equation}
In the diffusive regime, the binary can disrupt via high-speed encounters with perturbers of mass $M_p$, a process characterised by a time scale $t_{\rm ds}$ \citep{heggie75,bt} \begin{equation}
    t_{\rm ds} \simeq 1.9 {\rm ~Gyr} 
    \left(\frac{k_d}{0.002}\right)
    \left(\frac{\sigma}{40 {\rm ~km~s^{-1}}}\right)
    \left(\frac{M_1+M_2}{20\Ms}\right)
    \left(\frac{20\Ms}{M_p}\right)^{2}
    \left(\frac{10{\rm ~pc}^{-3}}{n_p}\right)
    \left(\frac{10{\rm ~AU}}{a}\right),
    \label{tds}
\end{equation}
where $k_d$ is a factor inferred from scattering experiments \citep{bahcall85} and $n_p$ represents the perturbers number density. However, if these encounters are sufficiently rare, the binary can undergo disruption secularly, due to the cumulative effect of all the weak interactions with cluster stars over. This process takes place over an evaporation time $t_e$ \citep{bt}:
\begin{equation}
    t_{\rm ev} \simeq 17.5 {\rm ~Gyr}
    \left(\frac{\sigma}{40 {\rm ~km~s^{-1}}}\right)
    \left(\frac{\rho_p}{100 ~\Ms ~{\rm pc}^{3}}\right)
    \left(\frac{a}{10{\rm ~AU}}\right)
    \left(\frac{{\rm log} \Lambda}{6.5}\right).
    \label{tevap}
\end{equation}

If the binary is hard, instead, an interaction with a passing-by star tends, on average, to harden the binary further \citep{heggie75}. As the binary hardens, the encounters become rarer but more violent, possibly resulting in the ejection of the binary over a time
\begin{equation}
    t_{\rm ej} = \left(\frac{k_{\rm hd}}{100}\right) \left(\frac{t_r}{0.13}\right),
    \label{tejec}
\end{equation}
where $t_r$ is the cluster relaxation time and $k_{\rm hd}$ is a parameter derived from scattering experiments \citep{Goodman93,Heggie03}. 


All the timescales listed above depend on the cluster properties and the average values of scattering parameters. However, our three-body simulations are a) uniquely defined by the cluster velocity dispersion, which is degenerate in the cluster mass and half-mass radius through Equation \ref{eq:2}, and b) do not take into account the disruption, evaporation, or ejection processes. To partly solve the degeneracy and provide a more reliable description of the possible NS-BH outcomes, we adopt the following treatment to determine whether the NS-BH is likely to survive further stellar encounters and eventually merge. 

For each NS-BH merger candidate we create a sample of 100 clusters with the same velocity dispersion but a half-mass radius selected from the observed distribution for GCs \citep{harris10} and NCs \citep{georgiev14,georgiev16}. For each cluster model, we create 100 different scattering parameters to obtain all the relevant timescales in Eqs. \ref{tau}-\ref{tejec}. First, we select the formation time through the following steps:
\begin{itemize}
    \item extract the $\tau$ value from the distribution reconstructed through MOCCA models;
    \item extract the cluster half-mass radius $r_h$ from the observed distribution of Galactic GCs and local NCs;
    \item combine $r_h$ and $\sigma$ (which is fixed for each model) to calculate the cluster mass $M_c$ and the corresponding relaxation time $t_r$ from Equation \ref{trelax};
\end{itemize}
for each $i$-th version of the same NS-BH merger candidate in a different environment, the formation time is thus uniquely defined as $t_{f,i} = \tau_i t_{r,i}$. This sampling is done for each value of the velocity dispersion explored here, i.e. $\sigma = 5,~15,~20,~35,~50,~100$ km s$^{-1}$.

To characterise the disruption processes, instead, we extract an impact parameter $b_i$ from a linear distribution $2bdb$, as expected from geometrical considerations, limited above by the cluster free mean path and below by 0.1 times the binary semimajor axis, i.e. sufficiently hard to pose a threat to the binary survival\footnote{We verified that decreasing further this limit does not impact the results.}. The mass of the perturber is extracted from a power-law mass function with slope $\alpha_{\rm MF} = -2.3$ limited in the range $M_p = (0.08-60)\Ms$. However, it must be noted that if the cluster core is dominated by heavy remnants, the mass function can be significantly steeper. For instance, if the cluster contains a BH subsystem, \cite{arcasedda18a} suggested that half of the mass inside the subsystem is contributed from stars and the remaining in BHs. If we assume that stellar BHs have masses in the range $(5-60) \Ms$ and that the overall mass function in the subsystem is described by a power-law, it is possible to show that inside the subsystem $\alpha_{\rm MF} \sim -1$. We found that a steeper mass function decreases the number of merger candidates by $12\%$. 

If the binary is soft, we check whether it is in the diffusive ($b_i > b_{\rm max}$) or in the catastrophic regime, calculating the corresponding disruption timescales. 

According to this statistical procedure, each NS-BH is characterised by 10,000 potential outcomes. We identify as potential mergers those fullfilling one of the following conditions:
\begin{itemize}
    \item $t_\gw + t_f < 14$ Gyr and $t_\gw < t_{\rm cat}$ if the binary is soft and in the catastrophic regime,
    \item $t_\gw + t_f < 14$ Gyr and $t_\gw < {\rm min} (t_{\rm ds},t_{\rm ev})$ if the binary is soft and in the diffusive regime,
    \item the binary is hard. 
\end{itemize}
If one of the conditions is fullfilled in at least $5\%$ of the interactions modelled for the same NS-BH candidate, we label it as a merger. The number of times the merging binary has been identified as hard ($N_{\rm hd}$) or soft ($N_{\rm sft}$) determines the binary status, which is labelled as ``hard'' if $N_{\rm hd}>N_{\rm sft}$ or ``soft'' otherwise. Similarly, we label the merger as in-cluster or ejected depending on the number of times that the binary has been identified as a candidate for ejection before merging or not.

We repeated the same procedure 10 times to verify the impact of randomization onto the calculation of the individual merger rate (Equation \ref{eq:1}) and we notice an overall variation of $2-4\%$ in the values quoted in Table \ref{tab:tab5}. 

In general, we find that the vast majority of NS-BH mergers in our models come from hard binaries ($>84-100\%$), with the fraction reaching the maximum in correspondence of lower-velocity dispersion clusters. Around $10\%$ of mergers in hard binaries and in clusters with $\sigma = 15-35$ km s$^{-1}$ are ejected from the parent cluster before the merger takes place.

\begin{table}
    \centering
    \begin{tabular}{cccccc}
    \hline
    \hline
configuration & $Z$ & $\sigma$ & $f_{\rm sft}$ & $f_{\rm hd,in}$ & $f_{\rm hd,ej}$ \\ 
              & [Z$_\odot$] & [km s$^{-1}$] & [\%] &[\%] &[\%]  \\
    \hline
\multirow{6}{*}{NSSTBH} & 0.0002 &5  & 0.0 & 100.0& 0.0 \\  
                        & 0.0002 &15 & 0.0 & 95.2 & 4.8 \\
                        & 0.0002 &20 & 0.0 & 90.3 & 9.7 \\
                        & 0.0002 &35 & 0.0 & 98.5 & 1.5 \\
                        & 0.0002 &50 & 1.7 & 98.3 & 0.0 \\
                        & 0.0002 &100& 15.5& 84.5 & 0.0 \\
    \hline
\multirow{6}{*}{BHSTNS} & 0.0002 &5  & 0.0 & 100.0& 0.0 \\
                        & 0.0002 &15 & 0.0 & 90.9 & 9.1 \\
                        & 0.0002 &20 & 0.0 & 95.5 & 4.6 \\
                        & 0.0002 &35 & 2.3 & 97.7 & 0.0 \\
                        & 0.0002 &50 & 7.3 & 91.0 & 1.8 \\
                        & 0.0002 &100& 5.5 & 94.5 & 0.0 \\
    \hline
\multirow{6}{*}{NSSTBH} & 0.02   &5  & 0.0 & 100.0& 0.0 \\
                        & 0.02   &15 & 4.0 & 88.0 & 8.0 \\
                        & 0.02   &20 & 0.0 & 88.1 & 11.9\\
                        & 0.02   &35 & 0.0 & 95.1 & 4.9 \\
                        & 0.02   &50 & 3.0 & 94.1 & 3.0 \\
                        & 0.02   &100& 10.6& 88.7 & 0.7 \\
    \hline
\multirow{6}{*}{BHSTNS} & 0.02   &5  & 0.0 & 100.0& 0.0 \\
                        & 0.02   &15 & 0.0 & 91.7 & 8.3 \\
                        & 0.02   &20 & 0.0 & 94.4 & 5.6 \\
                        & 0.02   &35 & 0.0 & 98.5 & 1.5 \\
                        & 0.02   &50 & 1.0 & 99.0 & 0.0 \\
                        & 0.02   &100& 4.4 & 95.6 & 0.0 \\
   \hline 
    \end{tabular}
    \caption{Col. 1: configuration. Col. 2: cluster metallicity. Col. 3: cluster velocity dispersion. Col. 4-6: percentage of mergers from soft binaries, from hard binaries merging inside the cluster, from hard binaries merging outside the cluster. }
    \label{tab:tab5}
\end{table}

\section{The role of the detector sensitivity}
\label{sec:app4}

\begin{figure}
    \centering
    \includegraphics[width=0.5\columnwidth]{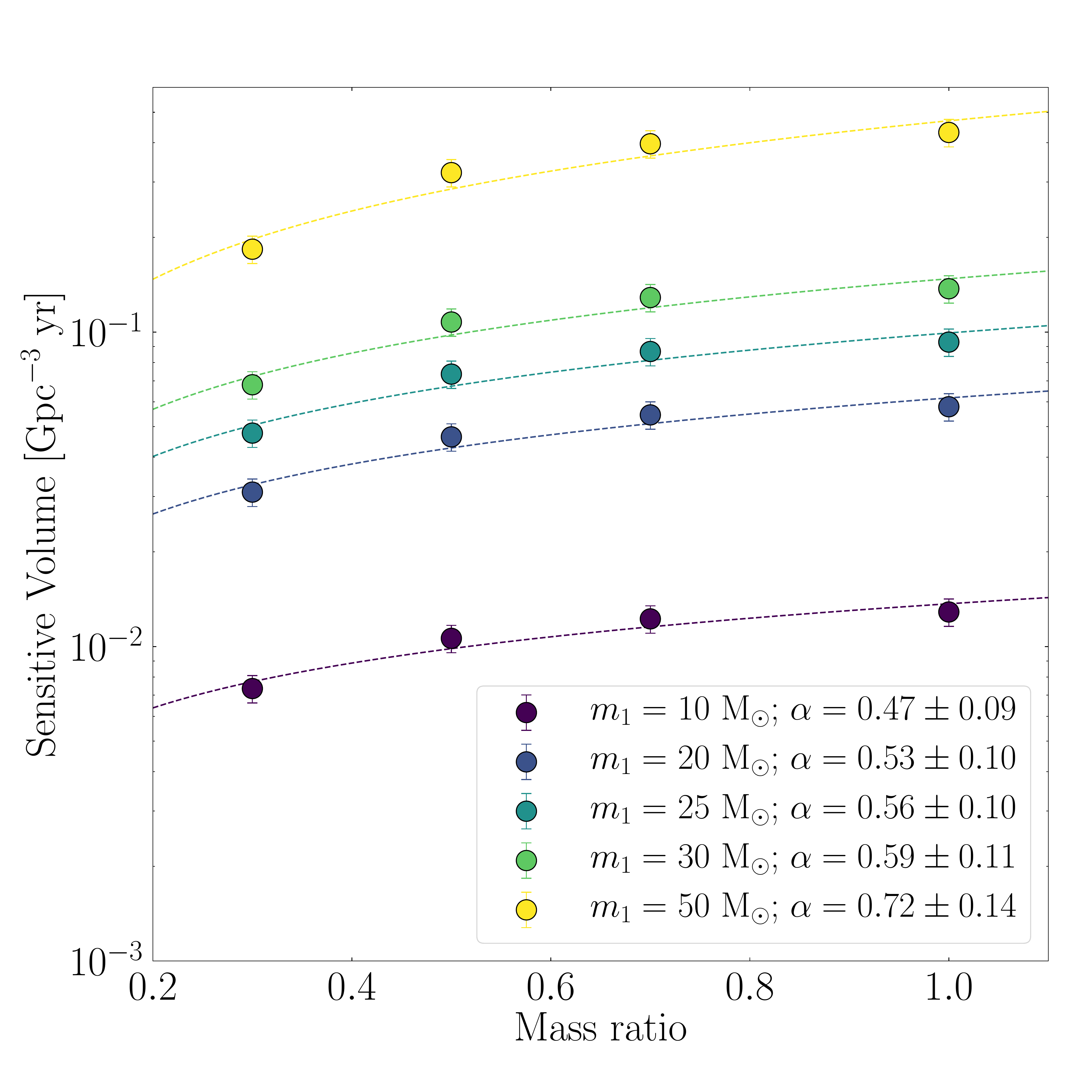}
    \caption{Sensitive redshifted spacetime volume, $VT$, of the LIGO detectors in observation runs O1 and O2 as a function of the mass ratio and for different values of the primary mass. The dotted lines represent least square fits of the data. The data point are extracted from \cite{fishbach17}.}
    \label{fig:fish}
\end{figure}

The detection of a GW source depends intrinsically on several parameters, such as the distance at which the merger took place, the direction of the wave hitting the detector, and the properties 
of the binary emitting GWs. Recently, \cite{fishbach17} have shown that the volume ($VT$) accessible to the LIGO detector scales with the mass of the primary through a power-law $VT\propto M_1^\alpha$, with $\alpha=2.2$, assuming $10<M_1/\Ms<100$ and at a fixed mass ratio, and increases at increasing the mass ratio if the primary mass is kept fixed (e.g. cfr. with their Figure 1). In order to take into account this observational bias in our analysis, we extract the data from \cite{fishbach17} Figure 1\footnote{We use the data extraction tool \url{https://apps.automeris.io/wpd/}.}. We associate to the extracted data a conservative error of $10\%$ and reconstruct the dependence between the volume and the mass ratio at fixed primary mass. As shown in Figure \ref{fig:fish}, we find, for primary masses in the range $10-50\Ms$, that the volume-mass ratio relation is well represented by a power-law $VT\propto q^\beta$ with a slope in the range $\beta = 0.47-0.72$. Given the mass of the primary in GW190814 we adopt $\beta = 0.5$ in the main analysis, although we test also the cases $\beta = 0.4$ and $\beta = 0.6$ for the sake of comparison.
Table \ref{tab:tab6} lists the probability $P_\lvk$ to find a GW190814-like merger in our database once this "volume weighting" procedure is taken into account. We find that varying the slope in the $VT-q$ relation causes a maximum variation in $P_\lvk$ of up to $\lesssim 10\%$. Due to this, in our calculations we assume a $P_\lvk = (22\pm 2) \% $ for solar metallicity systems and $P_\lvk = (4.3 \pm 0.4)\%$ for mergers with metal-poor progenitors. 

\begin{table}
    \centering
    \begin{tabular}{cccc}
    \hline\hline
        Z & \multicolumn{3}{c}{$P_\lvk$}\\
          &  $\beta = 0.4$ & $\beta = 0.5$ &  $\beta = 0.6$ \\
    \hline
        0.0002 & 4.04  & 4.32  & 4.63  \\ 
        0.02   & 23.2  & 21.8  & 22.1  \\ 
    \hline
    \end{tabular}
    \caption{ Col. 1: metallicity. Col. 2-4: percentage of models with GW190814-like mass and mass ratio assuming that the accessible volume scales with the mass ratio as $q^\beta$.}
    \label{tab:tab6}
\end{table}

\section{Comparison with similar works}
\label{sec:app3}

In this section we compare our inferred merger rate with recent results obtained through full $N$-body simulations of YCs \citep{rastello20, fragione20}. 

\begin{figure}
    \centering
    \includegraphics[width=0.6\columnwidth]{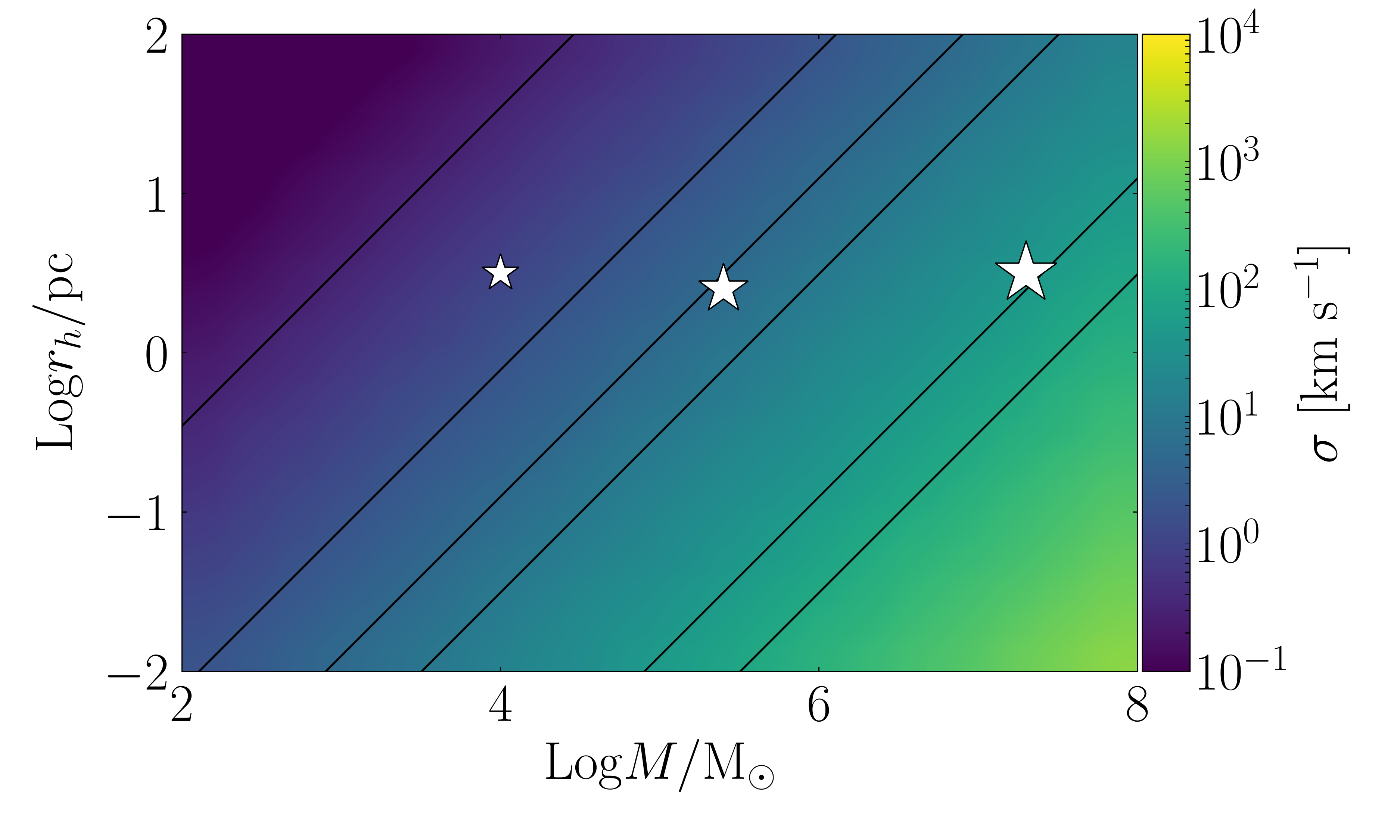}
    \caption{Surface map showing the cluster velocity dispersion as a function of mass and half-mass radius. From left to right, black lines identify clusters with a velocity dispersion of $0.3,~ 2,~ 5,~ 10,~ 50,~ 100$ km s$^{-1}$, respectively. The three black stars identify the typical values for Galactic YCs \citep{zwart10}, GCs \citep{harris10}, and the Galactic NC \citep{feldmeier14}}.
    \label{fig:clpr}
\end{figure}

To perform the comparison, we make use of Equation \ref{eq:2} to derive the mass, half-mass radius, or velocity dispersion of host clusters. For the sake of comparison, Figure \ref{fig:clpr} shows the link between these three parameters and highlights typical values for Galactic GCs, YCs, and the NC. 

\cite{rastello20} predict a global merger rate of $28\yrgpc$, but only $\sim 36-55\%$ of these mergers form dynamically, thus implying a {\it dynamical} merger rate of $\Gamma \simeq 14\yrgpc$. The simulations presented by \cite{rastello20} focus on star clusters with masses in the range $M = (0.3-1) \times 10^3\Ms$ and half-mass radii $R_h = 0.21-0.25$ pc, corresponding to a velocity dispersion $\sigma = 0.67-1.1$ km s$^{-1}$, according to our Equation \ref{eq:2}.

In this range of $\sigma$ values, our inferred individual merger rate is roughly $\Gamma_{\rm ind} \sim (3 \times 10^{-4} - 3\times 10^{-2}) $ Gyr$^{-1}$, depending on the metallicity (the larger the metallicity the lower the rate) and configuration (BHSTNS produces more mergers).

If we assume that \cite{rastello20} models represent the ``normal'' population of YCs in a MW-sized galaxy, we infer a merger rate of $\Gamma_{\rm YC} \sim 0.1-13\yrgpc$ (see Equation \ref{eq:YCrate} in the main paper), thus bracketing the value inferred from direct models. Note that the assumption that the density of YCs in the local Universe is 2.31 Mpc$^{-3}$ \citep[e.g. similar to that of globular clusters][]{rodriguez16,Ye2020,fragione20} leads to a merger rate of $\Gamma_{\rm YC} = (7\times 10^{-4} - 6.9\times 10^{-2}) \yrgpc$. 
In a more recent work, \cite{fragione20} explored the output of 65 $N$-body simulations of clusters with masses $M = 10^4-10^5\Ms$ and half-mass radii $R_h = 1-3$ pc \citep{banerjee20}. The authors find 15 NS-BH binaries formed over a 10 Gyr timescale, but none of them merging within a Hubble time. Adopting an average density for the cosmic YCs population of $\rho_{\rm YC} \sim 2.31$ Mpc$^{-3}$, the authors derive an upper limit on the NS-BH merger rate of $3\times10^{-3}\yrgpc$. 
For these models, the velocity dispersion inferred from Equation 2 in the main paper is $\sigma \simeq 1-5.6$ km s$^{-1}$. According to our calculations, the individual merger rate for these types of clusters is $\Gamma_{\rm ind} = (9.6\times 10^{-5} - 0.057)$ Gyr$^{-1}$, depending on the configuration and the metallicity. This implies a local Universe merger rate of $\Gamma = \rho \Gamma_{\rm ind} \sim (2.2\times 10^{-4} - 0.13) \yrgpc$, thus embracing predictions for both massive YCs \citep{fragione20} and globular clusters \citep[$0.055-5.5\yrgpc$, ][]{Ye2020}.

Despite our models rely upon several physically motivated key assumptions, they capture the essential elements of dynamical NS-BH formation, leading to merger rates in broad agreement with the results obtained with more detailed models. Our models thus provide a complementary view on the NS-BH mergers formation process, enabling the possibility to collect statistically significant sample of candidates that can be used to place constraints on the properties of the overall NS-BH merger population.

\bibliography{apssamp}{}

\newcommand{\noop}[1]{}
\begin{thebibliography}{}
\expandafter\ifx\csname natexlab\endcsname\relax\def\natexlab#1{#1}\fi
\providecommand{\url}[1]{\href{#1}{#1}}
\providecommand{\dodoi}[1]{doi:~\href{http://doi.org/#1}{\nolinkurl{#1}}}
\providecommand{\doeprint}[1]{\href{http://ascl.net/#1}{\nolinkurl{http://ascl.net/#1}}}
\providecommand{\doarXiv}[1]{\href{https://arxiv.org/abs/#1}{\nolinkurl{https://arxiv.org/abs/#1}}}

\bibitem[{{Abadie} {et~al.}(2010){Abadie}, {Abbott}, {Abbott}, {LIGO Scientific
  Collaboration}, \& {Virgo Collaboration}}]{abadie10}
{Abadie}, J., {Abbott}, B.~P., {Abbott}, R., {LIGO Scientific Collaboration},
  \& {Virgo Collaboration}. 2010, Classical and Quantum Gravity, 27, 173001,
  \dodoi{10.1088/0264-9381/27/17/173001}

\bibitem[{{Arca Sedda}(2020)}]{arcasedda20}
{Arca Sedda}, M. 2020, Communications Physics, 3, 43,
  \dodoi{10.1038/s42005-020-0310-x}

\bibitem[{{Arca Sedda} {et~al.}(2018){Arca Sedda}, {Askar}, \&
  {Giersz}}]{arcasedda18a}
{Arca Sedda}, M., {Askar}, A., \& {Giersz}, M. 2018, \mnras, 479, 4652,
  \dodoi{10.1093/mnras/sty1859}

\bibitem[{{Arca Sedda} \& {Benacquista}(2019)}]{arcasedda19b}
{Arca Sedda}, M., \& {Benacquista}, M. 2019, \mnras, 482, 2991,
  \dodoi{10.1093/mnras/sty2764}

\bibitem[{{Arca-Sedda} \& {Capuzzo-Dolcetta}(2019)}]{arcasedda19}
{Arca-Sedda}, M., \& {Capuzzo-Dolcetta}, R. 2019, \mnras, 483, 152,
  \dodoi{10.1093/mnras/sty3096}

\bibitem[{{Arca Sedda} {et~al.}(2020){Arca Sedda}, {Mapelli}, {Spera},
  {Benacquista}, \& {Giacobbo}}]{arcasedda20b}
{Arca Sedda}, M., {Mapelli}, M., {Spera}, M., {Benacquista}, M., \& {Giacobbo},
  N. 2020, \apj, 894, 133, \dodoi{10.3847/1538-4357/ab88b2}

\bibitem[{{Askar} {et~al.}(2017){Askar}, {Szkudlarek}, {Gondek-Rosi{\'n}ska},
  {Giersz}, \& {Bulik}}]{askar17}
{Askar}, A., {Szkudlarek}, M., {Gondek-Rosi{\'n}ska}, D., {Giersz}, M., \&
  {Bulik}, T. 2017, \mnras, 464, L36, \dodoi{10.1093/mnrasl/slw177}

\bibitem[{{Bahcall} {et~al.}(1985){Bahcall}, {Hut}, \& {Tremaine}}]{bahcall85}
{Bahcall}, J.~N., {Hut}, P., \& {Tremaine}, S. 1985, \apj, 290, 15,
  \dodoi{10.1086/162953}

\bibitem[{{Bailyn} {et~al.}(1998){Bailyn}, {Jain}, {Coppi}, \&
  {Orosz}}]{Bailyn98}
{Bailyn}, C.~D., {Jain}, R.~K., {Coppi}, P., \& {Orosz}, J.~A. 1998, \apj, 499,
  367, \dodoi{10.1086/305614}

\bibitem[{{Banerjee}(2020)}]{banerjee20}
{Banerjee}, S. 2020, arXiv e-prints, arXiv:2004.07382.
\newblock \doarXiv{2004.07382}

\bibitem[{{Belczynski} {et~al.}(2002){Belczynski}, {Kalogera}, \&
  {Bulik}}]{belczynski02}
{Belczynski}, K., {Kalogera}, V., \& {Bulik}, T. 2002, \apj, 572, 407,
  \dodoi{10.1086/340304}

\bibitem[{{Binney} \& {Tremaine}(2008)}]{bt}
{Binney}, J., \& {Tremaine}, S. 2008, {Galactic Dynamics: Second Edition}
  (Princeton University Press)

\bibitem[{{Breen} \& {Heggie}(2013)}]{breen13}
{Breen}, P.~G., \& {Heggie}, D.~C. 2013, \mnras, 432, 2779,
  \dodoi{10.1093/mnras/stt628}

\bibitem[{{Burrows} {et~al.}(2019){Burrows}, {Radice}, \&
  {Vartanyan}}]{burrows19}
{Burrows}, A., {Radice}, D., \& {Vartanyan}, D. 2019, \mnras, 485, 3153,
  \dodoi{10.1093/mnras/stz543}

\bibitem[{{Clausen} {et~al.}(2013){Clausen}, {Sigurdsson}, \&
  {Chernoff}}]{clausen13}
{Clausen}, D., {Sigurdsson}, S., \& {Chernoff}, D.~F. 2013, \mnras, 428, 3618,
  \dodoi{10.1093/mnras/sts295}

\bibitem[{{Dominik} {et~al.}(2012){Dominik}, {Belczynski}, {Fryer}, {Holz},
  {Berti}, {Bulik}, {Mand el}, \& {O'Shaughnessy}}]{dominik12}
{Dominik}, M., {Belczynski}, K., {Fryer}, C., {et~al.} 2012, \apj, 759, 52,
  \dodoi{10.1088/0004-637X/759/1/52}

\bibitem[{{Eldridge} {et~al.}(2017){Eldridge}, {Stanway}, {Xiao}, {McClelland
  }, {Taylor}, {Ng}, {Greis}, \& {Bray}}]{Eldridge17}
{Eldridge}, J.~J., {Stanway}, E.~R., {Xiao}, L., {et~al.} 2017, \pasa, 34,
  e058, \dodoi{10.1017/pasa.2017.51}

\bibitem[{{Feldmeier} {et~al.}(2014){Feldmeier}, {Neumayer}, {Seth},
  {Sch{\"o}del}, {L{\"u}tzgendorf}, {de Zeeuw}, {Kissler-Patig}, {Nishiyama},
  \& {Walcher}}]{feldmeier14}
{Feldmeier}, A., {Neumayer}, N., {Seth}, A., {et~al.} 2014, \aap, 570, A2,
  \dodoi{10.1051/0004-6361/201423777}

\bibitem[{{Fishbach} \& {Holz}(2017)}]{fishbach17}
{Fishbach}, M., \& {Holz}, D.~E. 2017, \apjl, 851, L25,
  \dodoi{10.3847/2041-8213/aa9bf6}

\bibitem[{{Fragione} \& {Banerjee}(2020)}]{fragione20}
{Fragione}, G., \& {Banerjee}, S. 2020, arXiv e-prints, arXiv:2006.06702.
\newblock \doarXiv{2006.06702}

\bibitem[{{Freire} {et~al.}(2008){Freire}, {Ransom}, {B{\'e}gin}, {Stairs},
  {Hessels}, {Frey}, \& {Camilo}}]{Freire08}
{Freire}, P. C.~C., {Ransom}, S.~M., {B{\'e}gin}, S., {et~al.} 2008, \apj, 675,
  670, \dodoi{10.1086/526338}

\bibitem[{{Fryer} {et~al.}(2012){Fryer}, {Belczynski}, {Wiktorowicz},
  {Dominik}, {Kalogera}, \& {Holz}}]{fryer12}
{Fryer}, C.~L., {Belczynski}, K., {Wiktorowicz}, G., {et~al.} 2012, \apj, 749,
  91, \dodoi{10.1088/0004-637X/749/1/91}

\bibitem[{{Georgiev} \& {B{\"o}ker}(2014)}]{georgiev14}
{Georgiev}, I.~Y., \& {B{\"o}ker}, T. 2014, \mnras, 441, 3570,
  \dodoi{10.1093/mnras/stu797}

\bibitem[{{Georgiev} {et~al.}(2016){Georgiev}, {B{\"o}ker}, {Leigh},
  {L{\"u}tzgendorf}, \& {Neumayer}}]{georgiev16}
{Georgiev}, I.~Y., {B{\"o}ker}, T., {Leigh}, N., {L{\"u}tzgendorf}, N., \&
  {Neumayer}, N. 2016, \mnras, 457, 2122, \dodoi{10.1093/mnras/stw093}

\bibitem[{{Giacobbo} \& {Mapelli}(2018)}]{giacobbo18}
{Giacobbo}, N., \& {Mapelli}, M. 2018, \mnras, 480, 2011,
  \dodoi{10.1093/mnras/sty1999}

\bibitem[{{Giesers} {et~al.}(2018){Giesers}, {Dreizler}, {Husser}, {Kamann},
  {Anglada Escud{\'e}}, {Brinchmann}, {Carollo}, {Roth}, {Weilbacher}, \&
  {Wisotzki}}]{giesers18}
{Giesers}, B., {Dreizler}, S., {Husser}, T.-O., {et~al.} 2018, \mnras, 475,
  L15, \dodoi{10.1093/mnrasl/slx203}

\bibitem[{{Goodman} \& {Hut}(1993)}]{Goodman93}
{Goodman}, J., \& {Hut}, P. 1993, \apj, 403, 271, \dodoi{10.1086/172200}

\bibitem[{{Harris}(2010)}]{harris10}
{Harris}, W.~E. 2010, arXiv e-prints, arXiv:1012.3224.
\newblock \doarXiv{1012.3224}

\bibitem[{{Heggie} \& {Hut}(2003)}]{Heggie03}
{Heggie}, D., \& {Hut}, P. 2003, Classical and Quantum Gravity, 20, 4504,
  \dodoi{10.1088/0264-9381/20/20/603}

\bibitem[{{Heggie}(1975)}]{heggie75}
{Heggie}, D.~C. 1975, \mnras, 173, 729, \dodoi{10.1093/mnras/173.3.729}

\bibitem[{{Hurley} {et~al.}(2002){Hurley}, {Tout}, \& {Pols}}]{hurley02}
{Hurley}, J.~R., {Tout}, C.~A., \& {Pols}, O.~R. 2002, \mnras, 329, 897,
  \dodoi{10.1046/j.1365-8711.2002.05038.x}

\bibitem[{{Jackson} {et~al.}(2020){Jackson}, {Jeffries}, {Wright}, {Rand ich},
  {Sacco}, {Pancino}, {Cantat-Gaudin}, {Gilmore}, {Vallenari}, {Bensby},
  {Bayo}, {Costado}, {Franciosini}, {Gonneau}, {Hourihane}, {Lewis}, {Monaco},
  {Morbidelli}, \& {Worley}}]{jackson20}
{Jackson}, R.~J., {Jeffries}, R.~D., {Wright}, N.~J., {et~al.} 2020, \mnras,
  \dodoi{10.1093/mnras/staa1749}

\bibitem[{{Kremer} {et~al.}(2020){Kremer}, {Ye}, {Rui}, {Weatherford},
  {Chatterjee}, {Fragione}, {Rodriguez}, {Spera}, \& {Rasio}}]{kremer20}
{Kremer}, K., {Ye}, C.~S., {Rui}, N.~Z., {et~al.} 2020, \apjs, 247, 48,
  \dodoi{10.3847/1538-4365/ab7919}

\bibitem[{{Kuhn} {et~al.}(2019){Kuhn}, {Hillenbrand}, {Sills}, {Feigelson}, \&
  {Getman}}]{kuhn19}
{Kuhn}, M.~A., {Hillenbrand}, L.~A., {Sills}, A., {Feigelson}, E.~D., \&
  {Getman}, K.~V. 2019, \apj, 870, 32, \dodoi{10.3847/1538-4357/aaef8c}

\bibitem[{{Liu} \& {Lai}(2020)}]{liu20}
{Liu}, B., \& {Lai}, D. 2020, arXiv e-prints, arXiv:2009.10068.
\newblock \doarXiv{2009.10068}

\bibitem[{{Lu} {et~al.}(2021){Lu}, {Beniamini}, \& {Bonnerot}}]{Lu21}
{Lu}, W., {Beniamini}, P., \& {Bonnerot}, C. 2021, \mnras, 500, 1817,
  \dodoi{10.1093/mnras/staa3372}

\bibitem[{{Mapelli} {et~al.}(2020){Mapelli}, {Spera}, {Montanari}, {Limongi},
  {Chieffi}, {Giacobbo}, {Bressan}, \& {Bouffanais}}]{mapelli19}
{Mapelli}, M., {Spera}, M., {Montanari}, E., {et~al.} 2020, \apj, 888, 76,
  \dodoi{10.3847/1538-4357/ab584d}

\bibitem[{{Marchant} {et~al.}(2017){Marchant}, {Langer}, {Podsiadlowski},
  {Tauris}, {de Mink}, {Mandel}, \& {Moriya}}]{marchant17}
{Marchant}, P., {Langer}, N., {Podsiadlowski}, P., {et~al.} 2017, \aap, 604,
  A55, \dodoi{10.1051/0004-6361/201630188}

\bibitem[{{McKernan} {et~al.}(2020){McKernan}, {Ford}, \&
  {O'Shaughnessy}}]{mckernan20}
{McKernan}, B., {Ford}, K.~E.~S., \& {O'Shaughnessy}, R. 2020, arXiv e-prints,
  arXiv:2002.00046.
\newblock \doarXiv{2002.00046}

\bibitem[{{Mikkola} \& {Merritt}(2008)}]{mikkola08}
{Mikkola}, S., \& {Merritt}, D. 2008, \aj, 135, 2398,
  \dodoi{10.1088/0004-6256/135/6/2398}

\bibitem[{{Mikkola} \& {Tanikawa}(1999)}]{mikkola99}
{Mikkola}, S., \& {Tanikawa}, K. 1999, \mnras, 310, 745,
  \dodoi{10.1046/j.1365-8711.1999.02982.x}

\bibitem[{{Morscher} {et~al.}(2015){Morscher}, {Pattabiraman}, {Rodriguez},
  {Rasio}, \& {Umbreit}}]{morscher15}
{Morscher}, M., {Pattabiraman}, B., {Rodriguez}, C., {Rasio}, F.~A., \&
  {Umbreit}, S. 2015, \apj, 800, 9, \dodoi{10.1088/0004-637X/800/1/9}

\bibitem[{{{\"O}zel} {et~al.}(2012){{\"O}zel}, {Psaltis}, {Narayan}, \& {Santos
  Villarreal}}]{Ozel12}
{{\"O}zel}, F., {Psaltis}, D., {Narayan}, R., \& {Santos Villarreal}, A. 2012,
  \apj, 757, 55, \dodoi{10.1088/0004-637X/757/1/55}

\bibitem[{Peters(1964)}]{peters64}
Peters, P.~C. 1964, Phys. Rev., 136, B1224, \dodoi{10.1103/PhysRev.136.B1224}

\bibitem[{{Piskunov} {et~al.}(2006){Piskunov}, {Kharchenko}, {R{\"o}ser},
  {Schilbach}, \& {Scholz}}]{piskunov06}
{Piskunov}, A.~E., {Kharchenko}, N.~V., {R{\"o}ser}, S., {Schilbach}, E., \&
  {Scholz}, R.~D. 2006, \aap, 445, 545, \dodoi{10.1051/0004-6361:20053764}

\bibitem[{{Piskunov} {et~al.}(2007){Piskunov}, {Schilbach}, {Kharchenko},
  {R{\"o}ser}, \& {Scholz}}]{piskunov07}
{Piskunov}, A.~E., {Schilbach}, E., {Kharchenko}, N.~V., {R{\"o}ser}, S., \&
  {Scholz}, R.~D. 2007, \aap, 468, 151, \dodoi{10.1051/0004-6361:20077073}

\bibitem[{{Podsiadlowski} {et~al.}(2004){Podsiadlowski}, {Langer},
  {Poelarends}, {Rappaport}, {Heger}, \& {Pfahl}}]{podsiadlowski04}
{Podsiadlowski}, P., {Langer}, N., {Poelarends}, A.~J.~T., {et~al.} 2004, \apj,
  612, 1044, \dodoi{10.1086/421713}

\bibitem[{{Portegies Zwart} {et~al.}(2010){Portegies Zwart}, {McMillan}, \&
  {Gieles}}]{zwart10}
{Portegies Zwart}, S.~F., {McMillan}, S. L.~W., \& {Gieles}, M. 2010, \araa,
  48, 431, \dodoi{10.1146/annurev-astro-081309-130834}

\bibitem[{{Rastello} {et~al.}(2020){Rastello}, {Mapelli}, {Di Carlo},
  {Giacobbo}, {Santoliquido}, {Spera}, \& {Ballone}}]{rastello20}
{Rastello}, S., {Mapelli}, M., {Di Carlo}, U.~N., {et~al.} 2020, arXiv
  e-prints, arXiv:2003.02277.
\newblock \doarXiv{2003.02277}

\bibitem[{{Rodriguez} {et~al.}(2018){Rodriguez}, {Amaro-Seoane}, {Chatterjee},
  {Kremer}, {Rasio}, {Samsing}, {Ye}, \& {Zevin}}]{rodriguez18}
{Rodriguez}, C.~L., {Amaro-Seoane}, P., {Chatterjee}, S., {et~al.} 2018, \prd,
  98, 123005, \dodoi{10.1103/PhysRevD.98.123005}

\bibitem[{{Rodriguez} {et~al.}(2016){Rodriguez}, {Chatterjee}, \&
  {Rasio}}]{rodriguez16}
{Rodriguez}, C.~L., {Chatterjee}, S., \& {Rasio}, F.~A. 2016, \prd, 93, 084029,
  \dodoi{10.1103/PhysRevD.93.084029}

\bibitem[{{Safarzadeh} \& {Loeb}(2020)}]{safarzadeh20}
{Safarzadeh}, M., \& {Loeb}, A. 2020, \apjl, 899, L15,
  \dodoi{10.3847/2041-8213/aba9df}

\bibitem[{{Sigurdsson} \& {Phinney}(1993)}]{sigurdsson93}
{Sigurdsson}, S., \& {Phinney}, E.~S. 1993, \apj, 415, 631,
  \dodoi{10.1086/173190}

\bibitem[{{Soubiran} {et~al.}(2018){Soubiran}, {Cantat-Gaudin},
  {Romero-G{\'o}mez}, {Casamiquela}, {Jordi}, {Vallenari}, {Antoja},
  {Balaguer-N{\'u}{\~n}ez}, {Bossini}, {Bragaglia}, {Carrera}, {Castro-Ginard},
  {Figueras}, {Heiter}, {Katz}, {Krone-Martins}, {Le Campion}, {Moitinho}, \&
  {Sordo}}]{soubiran18}
{Soubiran}, C., {Cantat-Gaudin}, T., {Romero-G{\'o}mez}, M., {et~al.} 2018,
  \aap, 619, A155, \dodoi{10.1051/0004-6361/201834020}

\bibitem[{{Spera} \& {Mapelli}(2017)}]{spera17}
{Spera}, M., \& {Mapelli}, M. 2017, \mnras, 470, 4739,
  \dodoi{10.1093/mnras/stx1576}

\bibitem[{{Spera} {et~al.}(2019){Spera}, {Mapelli}, {Giacobbo}, {Trani},
  {Bressan}, \& {Costa}}]{spera19}
{Spera}, M., {Mapelli}, M., {Giacobbo}, N., {et~al.} 2019, \mnras, 485, 889,
  \dodoi{10.1093/mnras/stz359}

\bibitem[{{The LIGO Scientific Collaboration} {et~al.}(2020){The LIGO
  Scientific Collaboration}, {the Virgo Collaboration}, \& et~al}]{190814}
{The LIGO Scientific Collaboration}, {the Virgo Collaboration}, \& et~al. 2020,
  arXiv e-prints, arXiv:2006.12611.
\newblock \doarXiv{2006.12611}

\bibitem[{{Thompson} {et~al.}(2019){Thompson}, {Kochanek}, {Stanek}, {Badenes},
  {Post}, {Jayasinghe}, {Latham}, {Bieryla}, {Esquerdo}, {Berlind}, {Calkins},
  {Tayar}, {Lindegren}, {Johnson}, {Holoien}, {Auchettl}, \&
  {Covey}}]{Thompson19}
{Thompson}, T.~A., {Kochanek}, C.~S., {Stanek}, K.~Z., {et~al.} 2019, Science,
  366, 637, \dodoi{10.1126/science.aau4005}

\bibitem[{{Tsokaros} {et~al.}(2020){Tsokaros}, {Ruiz}, \& {Shapiro}}]{ruiz20}
{Tsokaros}, A., {Ruiz}, M., \& {Shapiro}, S.~L. 2020, arXiv e-prints,
  arXiv:2007.05526.
\newblock \doarXiv{2007.05526}

\bibitem[{{Wyrzykowski} \& {Mandel}(2020)}]{Wyrzykowski20}
{Wyrzykowski}, {\L}., \& {Mandel}, I. 2020, \aap, 636, A20,
  \dodoi{10.1051/0004-6361/201935842}

\bibitem[{{Yang} {et~al.}(2019){Yang}, {Bartos}, {Haiman}, {Kocsis},
  {M{\'a}rka}, {Stone}, \& {M{\'a}rka}}]{yang20}
{Yang}, Y., {Bartos}, I., {Haiman}, Z., {et~al.} 2019, \apj, 876, 122,
  \dodoi{10.3847/1538-4357/ab16e3}

\bibitem[{{Yang} {et~al.}(2020){Yang}, {Gayathri}, {Bartos}, {Haiman},
  {Safarzadeh}, \& {Tagawa}}]{yang20b}
{Yang}, Y., {Gayathri}, V., {Bartos}, I., {et~al.} 2020, \apjl, 901, L34,
  \dodoi{10.3847/2041-8213/abb940}

\bibitem[{{Ye} {et~al.}(2020){Ye}, {Fong}, {Kremer}, {Rodriguez}, {Chatterjee},
  {Fragione}, \& {Rasio}}]{Ye2020}
{Ye}, C.~S., {Fong}, W.-f., {Kremer}, K., {et~al.} 2020, \apjl, 888, L10,
  \dodoi{10.3847/2041-8213/ab5dc5}

\bibitem[{{Zevin} {et~al.}(2020){Zevin}, {Spera}, {Berry}, \&
  {Kalogera}}]{zevin20}
{Zevin}, M., {Spera}, M., {Berry}, C. P.~L., \& {Kalogera}, V. 2020, arXiv
  e-prints, arXiv:2006.14573.
\newblock \doarXiv{2006.14573}

\bibitem[{{Ziosi} {et~al.}(2014){Ziosi}, {Mapelli}, {Branchesi}, \&
  {Tormen}}]{ziosi14}
{Ziosi}, B.~M., {Mapelli}, M., {Branchesi}, M., \& {Tormen}, G. 2014, \mnras,
  441, 3703, \dodoi{10.1093/mnras/stu824}

\end{thebibliography}
\bibliographystyle{aasjournal}

\end{document}